\documentclass[10pt]{iopart}
\usepackage{graphicx,bm}
\eqnobysec

\newcommand{\beq}{\begin{equation}}
\newcommand{\beqa}{\begin{eqnarray}}
\newcommand{\eeq}{\end{equation}}
\newcommand{\eeqa}{\end{eqnarray}}
\renewcommand{\Re}{\mathop{\rm Re}}

\newcommand{\abs}[1]{\vert#1\vert}
\renewcommand{\bar}[1]{{\overline{#1}}}
\newcommand{\bigmean}[1]{\left\langle#1\right\rangle}
\newcommand{\dd}{{\rm d}}
\newcommand{\h}[1]{{\widehat{#1}}}
\newcommand{\ii}{{\rm i}}
\newcommand{\lam}{\lambda}
\newcommand{\mean}[1]{\langle#1\rangle}
\newcommand{\prob}[1]{{{\rm Prob}\{#1\}}}
\renewcommand{\vec}[1]{{\bm#1}}
\newcommand{\lap}[1]{\mathrel{\mathop{\cal L}\limits_{#1}^{}}}

\begin{document}

\title{On sequences of records generated by planar random walks}

\author{Claude Godr\`eche and Jean-Marc Luck}

\address{Universit\'e Paris-Saclay, CNRS, CEA, Institut de Physique Th\'eorique,
91191~Gif-sur-Yvette, France}

\begin{abstract}
We investigate the statistics of three kinds of records associated with planar random walks,
namely diagonal, simultaneous and radial records.
The mean numbers of these records grow as universal power laws of time,
with respective exponents 1/4, 1/3 and 1/2.
The study of diagonal and simultaneous records relies on the underlying renewal structure of
the successive hitting times and locations of translated copies of a fixed target.
In this sense, this work represents a two-dimensional extension of
the analysis made by Feller
of ladder points, i.e., records for one-dimensional random walks.
This approach yields a variety of analytical asymptotic results,
including the full statistics of the numbers of diagonal and simultaneous records,
the joint law of the epoch and location of the current diagonal record
and the angular distribution of the current simultaneous record.
The sequence of radial records cannot be constructed in terms of a renewal process.
In spite of this,
their mean number is shown to grow with a super-universal square-root law
for isotropic random walks in any spatial dimension.
Their full distribution is also obtained.
Higher-dimensional diagonal and simultaneous records are also briefly discussed.
\end{abstract}

\eads{\mailto{claude.godreche@ipht.fr},\mailto{jean-marc.luck@ipht.fr}}

\maketitle

\section{Introduction}
\label{intro}

An observation in a time series is called an (upper) record if it is larger than all previous observations in the series.
The simplest case is when these observations are independent and identically distributed
(iid)~\cite{chandler,foster,ren2,gli,arn,nev,bun}.
However, in many instances observations are not iid random variables.
Consider for example the sequence of positions of a one-dimensional random walker.
The points where the position reaches a record value, that is, where it exceeds all previously attained values,
are called the \textit{ladder points}, in the terminology of Feller~\cite{feller1,feller2}.
Ladder points are important because sections between them are probabilistic replicas of each other.
This renewal structure makes the study of records for one-dimensional random walks amenable to exact analysis
(see~\cite[chapters XII and XVIII]{feller2}).

More recently, the statistics of records has found applications in a variety of complex physical systems
such as disordered systems, aging systems, complex networks, biological and geophysical processes, to name but a few.
We refer the reader to the reviews~\cite{wer,gmsrev} for panoramas of recent applications
of the theory of records in statistical physics.
In particular the topic of records for one-dimensional random walks has been
revisited and enriched in a series of papers in the past two decades
(see~\cite{gmsrev} and the references therein).

In this paper we investigate the statistics of various sequences of records associated with planar random walks.
The present work is therefore a natural extension of Feller's pioneering studies recalled above,
and of the subsequent investigations on the same topic reviewed in~\cite{gmsrev}.
To the best of our knowledge, this higher-dimensional case is entirely novel,
with the exception of a numerical study of radial records in one, two and three dimensions~\cite{edery}.

Consider for definiteness the simple random walk on the square lattice,
also referred to as the planar Polya walk~\cite{Polya}.
Let
\beq\label{eq:Rt}
\vec R_t=(X_t,Y_t)
\eeq
denote the random position of a walker launched from the origin after $t$ discrete time steps.
At variance with the one-dimensional situation,
a great many different kinds of records can be attached to a planar random walk.
The simplest situation consists in extracting a one-dimensional time series
from the walk by monitoring either one coordinate of the walker, or its radius, its polar angle,
and more generally any scalar function of its two coordinates.
In Section~\ref{rad} we investigate the example of radial records,
i.e., records of the radius $R_t=\abs{\vec R_t}$ of the walker, such that
\begin{itemize}
\item[(R)] there is a \textit{radial record} at time $t$ if:\\
$R_t>R_s$ for all $s=0,\dots,t-1$.
\end{itemize}
One may also think of constructions involving the coordinates $X_t$ and $Y_t$ in a more intricate fashion.
In this work we consider two examples which possess a renewal structure.
Simultaneous records, to be studied in Section~\ref{simul},
are the simultaneous occurrences of records for both coordinates of the random walk, such that
\begin{itemize}
\item[(S)] there is a \textit{simultaneous record} at time $t$ if:\\
$X_t>X_s$ and $Y_t>Y_s$ for all $s=0,\dots,t-1$.
\end{itemize}
Diagonal records, to be investigated in Section~\ref{diag},
are the records of the intersections of the random walk with the main diagonal, such that
\begin{itemize}
\item[(D)] there is a \textit{diagonal record} at time $t$ if:\\
$X_t=Y_t$ and $X_t>X_s$ for all those $s=0,\dots,t-1$ such that $X_s=Y_s$.
\end{itemize}

In numerical studies we shall consider in parallel two different kinds of planar random walks
with discrete steps of unit length,
namely the Polya walk, i.e., the simple random walk on the square lattice,
and the Pearson walk~\cite{Pearson}, whose steps have unit length and uniform random orientations.
In both cases a walk of $t$ steps issued from the origin obeys
$\mean{\vec R_t^2}=\mean{X_t^2+Y_t^2}=t$.
The diffusion coefficient $D$, such that $\mean{X_t^2}=\mean{Y_t^2}=2Dt$, is therefore
\beq
D=\frac14
\label{d14}
\eeq
for both walks.
References~\cite{spitzer,weiss,hughes} provide some general background on random walks.
The above definitions of radial and simultaneous records
are extended to off-lattice walks in a straightforward way,
whereas the definition of diagonal records has to be slightly adapted.
For the Pearson walk, and any other kind of off-lattice walk,
each step crossing the diagonal generates an intersection point,
to be determined by linear interpolation.
Diagonal records are the records of those intersection points.

\begin{figure}[!ht]
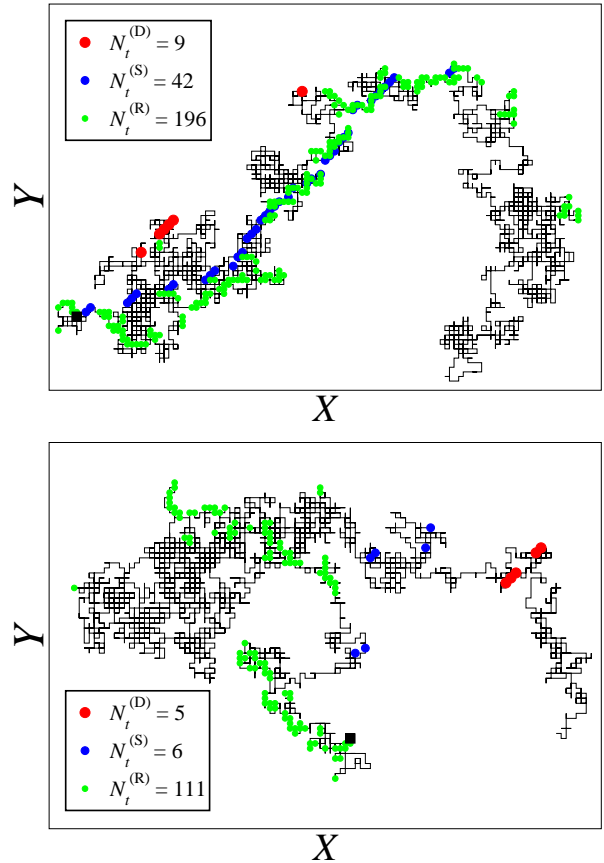

\begin{center}
\includegraphics[angle=0,width=.6\linewidth,clip=true]{exemple1.eps}
\vskip 6pt
\includegraphics[angle=0,width=.6\linewidth,clip=true]{exemple2.eps}
\caption{\small
Two Polya walks with $t=5000$ steps.
Black square: origin.
Red symbols: the $N_t^{({\rm D})}$ diagonal records.
Blue symbols: the~$N_t^{({\rm S})}$ simultaneous records.
Green symbols: the $N_t^{({\rm R})}$ radial records.
The legend gives the numbers of records of each kind.}
\label{exemples}
\end{center}
\end{figure}

Figure~\ref{exemples} shows two Polya walks of 5000 steps
with their diagonal, simultaneous, and radial records.
The numbers and patterns of the three kinds of records under consideration
look very different from each other.
The setup of this paper parallels these illustrations.
Diagonal, simultaneous, and radial records of planar walks are successively
investigated in Sections~\ref{diag},~\ref{simul}, and~\ref{rad}.
The mean numbers of records of each kind will be shown
to grow as different powers of time $t$, i.e.,
\beq
N_t^{({\rm D})}\sim t^{1/4},\quad
N_t^{({\rm S})}\sim t^{1/3},\quad
N_t^{({\rm R})}\sim t^{1/2}.
\label{123}
\eeq
Diagonal and simultaneous records are amenable to an exact analysis,
because they amount to renewal processes,
whereas our study of radial records is partly heuristic.
Section~\ref{disc} contains a summary of our main findings
and a discussion focussed onto records for isotropic random walks in higher spatial dimensions.
Three appendices are devoted to more technical matters.

\section{Diagonal records}
\label{diag}

\subsection{Recursive construction}
\label{diagrecur}

\begin{figure}[!ht]
\begin{center}
\includegraphics[angle=0,width=.4\linewidth,clip=true]{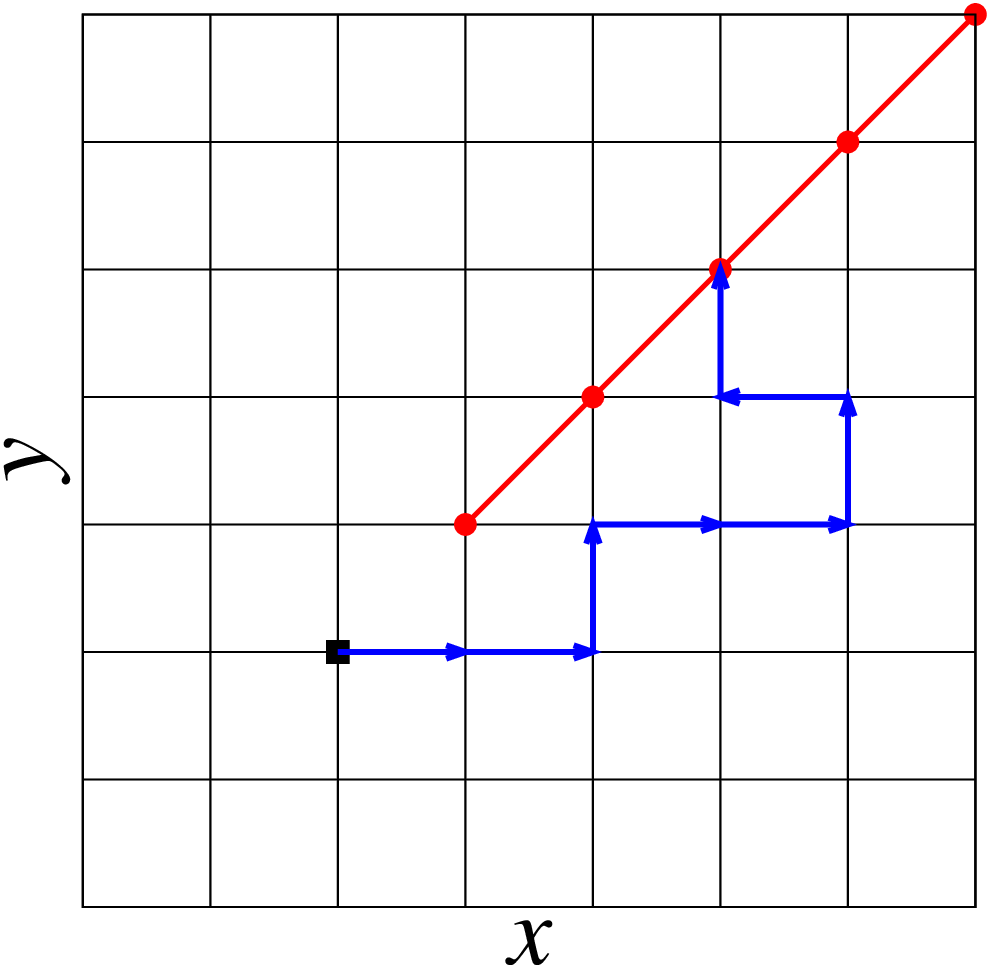}
\caption{\small
First step of recursive construction of diagonal records of a Polya walk.
Black square: origin.
Red: target (diagonal).
Blue: Polya walk.}
\label{cstrdiag}
\end{center}
\end{figure}

This section is devoted to the statistics of diagonal records,
shown as red symbols in Figure~\ref{exemples}.
Let us consider the Polya walk for the time being.
This process is Markovian and invariant under lattice translations.
Diagonal records therefore admit a recursive description,
whose first step is illustrated in Figure~\ref{cstrdiag}.
The target is the set of diagonal points of the lattice,
starting with $(1,1)$ and marked in red.
The first diagonal record corresponds to the first hitting of the target
by a walk issued from the origin.
On the example, the walk makes $\tau_1=8$ steps
before it hits the target at abscissa $x_1=3$.
The first diagonal record therefore occurs at time $\tau_1=8$ and abscissa $x_1=3$.

The second record can be constructed by considering
the location $(x_1,x_1)=(3,3)$ of the first record as a new origin.
The walk issued from that origin makes $\tau_2$ steps
before it hits the diagonal at abscissa $x_2$, and so on.
Such a recursive construction defines a renewal process.
References~\cite{feller2,dynkin,cox,cox-miller} provide overviews of classical renewal theory.
As recalled in Section~\ref{intro},
renewal theory has already been used in the study
of records associated to random walks in one dimension~\cite{feller2}
(see also~\cite{gmsrev} and the references therein).
The present construction is a higher-dimensional extension of these earlier works.

The $n$th diagonal record takes place at time $T_{(n)}$ and at abscissa $X_{(n)}$,
i.e., at the lattice point $\vec R_{(n)}=(X_{(n)},X_{(n)})$, where
\beqa
T_{(n)}&=&\tau_1+\cdots+\tau_n,
\nonumber\\
X_{(n)}&=&x_1+\cdots+x_n.
\label{partials}
\eeqa

Temporal and spatial increments $(\tau_n,x_n)$
are iid couples of integer random variables,
distributed according to the joint law $p(\tau,x)$
of the hitting time~$\tau$ and of the abscissa $x$ of the hitting point along the diagonal,
for a random walk starting at the origin, as shown in Figure~\ref{cstrdiag}.
We notice that $\tau$ is even.
To our knowledge, no expression for the exact distribution $p(\tau,x)$ is known.
It is worth stressing that the combinatorics of constrained lattice walks
is still a topical subject in the community of discrete mathematics,
as testified by the recent works~\cite{MBM,RT} on planar lattice walks avoiding a quadrant,
and by the many references therein.
The problem studied in those references
is in some sense related to the statistics of simultaneous records,
to be studied in Section~\ref{simul}.

Throughout the following we are mostly interested in asymptotic results
on the statistics of records in the scaling regime of large times and distances.
For that purpose, it is sufficient to know the asymptotic behavior
of the joint distribution $p(\tau,x)$ when both variables are large.
This asymptotic form can be derived from the continuum diffusion theory.
A self-contained presentation of this approach is given in Section~\ref{brown}.
Diagonal and simultaneous records, studied in Sections~\ref{diag} and~\ref{simul},
are related to the survival of a Brownian particle
in wedges of respective angles $\alpha=2\pi$ (cut plane) and $\alpha=3\pi/2$
(complement of a quadrant).

\subsection{Survival of a Brownian particle in a wedge}
\label{brown}

\begin{figure}[!ht]
\begin{center}
\includegraphics[angle=0,width=.5\linewidth,clip=true]{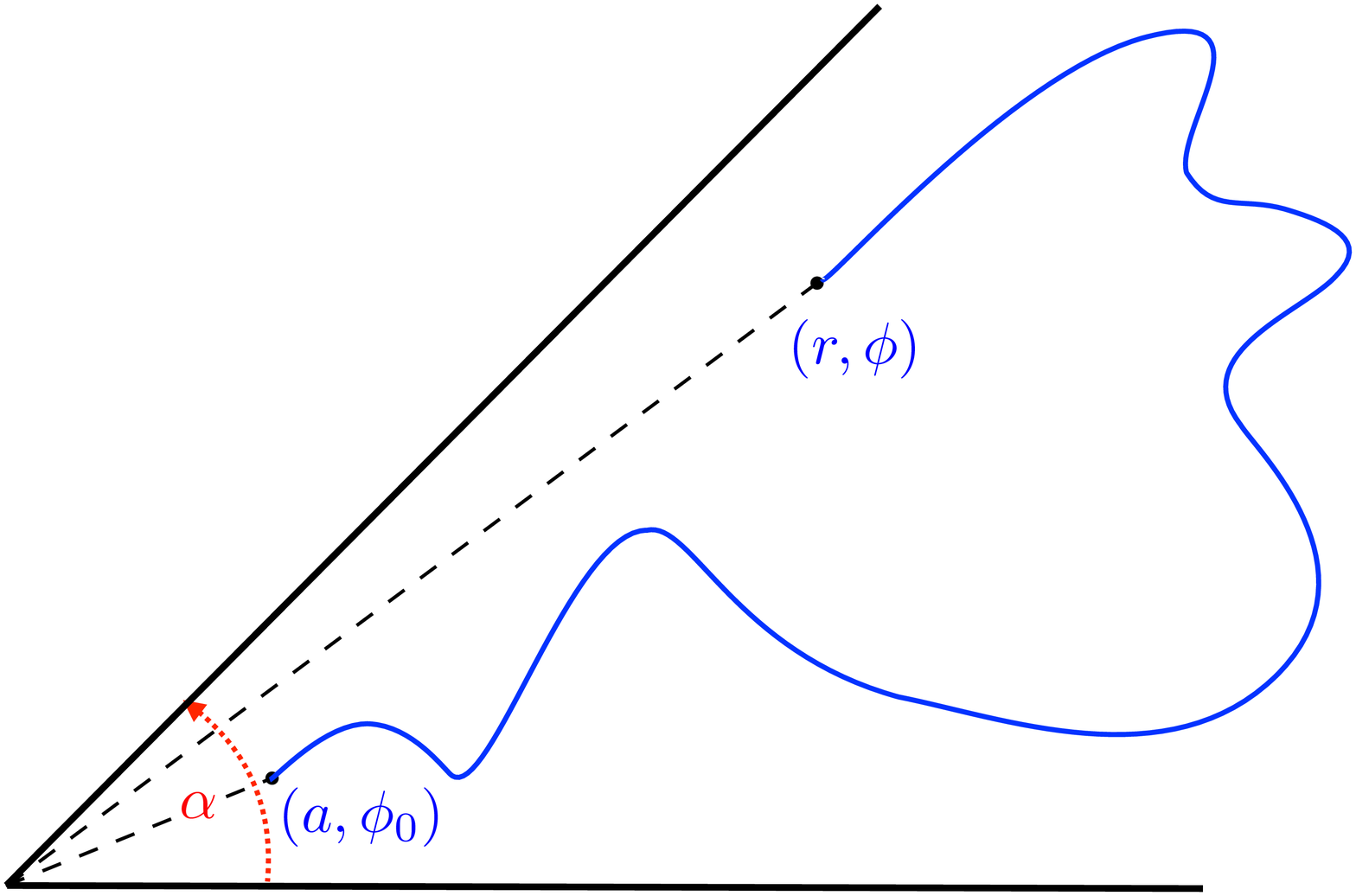}
\caption{\small
Geometric setting of survival of a Brownian particle in a wedge of angle $\alpha$.}
\label{secteur}
\end{center}
\end{figure}

The problem of the survival of a Brownian particle in a wedge of arbitrary angle~$\alpha$,
illustrated in Figure~\ref{secteur},
has been considered by Sommerfeld~\cite{sommerfeld}
and revisited many times in the modern era~\cite{redbook,CDB,laghol,DY,BKC,BKM,CBM}.
This section presents a comprehensive analysis of the problem,
with an emphasis on the distribution of the hitting time $\tau$ and hitting distance $x$,
corresponding to the event
where the particle hits the wedge boundary for the first time.

In polar coordinates $r>0$ and $0<\phi<\alpha$,
the probability density $G(r,\phi,a,\phi_0,t)$
for a Brownian particle issued from $(a,\phi_0)$ at time $t=0$
to have survived (i.e., stayed in the wedge) until time $t$
and be at point $(r,\phi)$ at time $t$ is referred to as the Green's function of the problem.
It obeys the diffusion equation
\beq
\frac{\partial G}{\partial t}=D\Delta G,
\label{diffeq}
\eeq
with initial value $G(r,\phi,a,\phi_0,0)=\delta(r-a)\delta(\phi-\phi_0)/a$
(as the integration measure is $r\,\dd r\,\dd\phi$),
and Dirichlet (i.e., absorbing) boundary conditions
$G(r,0,a,\phi_0,t)=G(r,\alpha,a,\phi_0,t)=0$ along both sides of the wedge.

The exact solution to~(\ref{diffeq}) reads (see e.g.~\cite{CDB,laghol})
\nopagebreak
\beqa
G(r,\phi,a,\phi_0,t)&=&\frac{\e^{-(a^2+r^2)/(4Dt)}}{\alpha Dt}
\nonumber\\
&\times&
\sum_{m=1}^\infty\sin(2m\theta\phi_0)\sin(2m\theta\phi)
\,I_{2m\theta}\!\left(\frac{ar}{2Dt}\right),
\label{gseries}
\eeqa
where $I_\nu$ is the modified Bessel function, and
\beq
\theta=\frac{\pi}{2\alpha}.
\label{thetadef}
\eeq

Let us fix once for all the starting point of the particle
at some microscopic distance $a$ from the origin,
on the symmetry axis of the wedge $(\phi_0=\alpha/2)$.
In the regime of interest, where both time $t$ and distance $r$ are macroscopically large,
the first term $(m=1)$ dominates the sum in~(\ref{gseries}).
Using the behavior $I_\nu(z)\approx(z/2)^\nu/\Gamma(\nu+1)$ as $z\to0$, with $\nu=2\theta$,
and expressing $\alpha$ in terms of $\theta$,
(\ref{gseries}) simplifies to
\beq
G(r,\phi,t)\approx\frac{\sin(2\theta\phi)}{\pi\Gamma(2\theta)Dt}
\left(\frac{ar}{4Dt}\right)^{2\theta}\e^{-r^2/(4Dt)}.
\label{gsimple}
\eeq

The first quantity of interest is the survival probability $S(t)$,
i.e., the probability that the particle has stayed in the wedge until time $t$.
This reads
\beq
S(t)=\int_0^\infty r\,\dd r\int_0^\alpha\dd\phi\,G(r,\phi,t).
\eeq
Integrating~(\ref{gsimple}) yields
\beq
S(t)\approx\frac{c}{\theta\,t^\theta},
\label{sasy}
\eeq
where
\beq
c=\frac{2\Gamma(\theta+1)}{\pi\Gamma(2\theta)}\left(\frac{a^2}{4D}\right)^\theta
\label{cbrown}
\eeq
is dubbed the tail parameter.
We have thus recovered the well-known result~\cite{redbook,CDB,laghol,DY,BKC,BKM,CBM}
that the survival probability falls off as a power law,
whose exponent~$\theta$ depends continuously on the wedge angle $\alpha$
according to~(\ref{thetadef}).
We have also determined the dependence of the tail parameter $c$ on the initial point,
within the continuum diffusion theory.

The distribution $f_\tau(\tau)$\footnote{Throughout this paper
$f_x(.)$ denotes the probability density of the continuous random variable~$x$.
For short this quantity is referred to as the distribution of $x$.
Similar notations are consistently used for multivariate and/or conditional probability densities.}
of the random hitting time $\tau$
where the particle hits the boundary of the wedge is simply related
to the survival probability $S(t)$.
This quantity is nothing but the probability that $\tau$ is larger than $t$:
\beq
S(t)=\int_t^\infty f_\tau(\tau)\dd\tau.
\label{sftau}
\eeq
By differentiating~(\ref{sasy}) with respect to $t$,
we obtain that the probability density of $\tau$ has a power-law tail of the form
\beq
f_\tau(\tau)\approx\frac{c}{\tau^{\theta+1}}.
\label{ftau}
\eeq

The full joint distribution $f_{\tau,r}(\tau,r)$ of the hitting time $\tau$
and of the distance $r$ between the hitting point and the origin
is given by the sum of the fluxes $\vec{n}\cdot\vec{J}$ of the probability current
$\vec{J}=-D\vec{\nabla}G$ at both points $(r,0)$ and $(r,\alpha)$
of the wedge boundary,
where $\vec{n}$ denote the corresponding internal normal vectors.
This reads
\beq
f_{\tau,r}(\tau,r)=\frac{D}{r}\left(
\left.\frac{\partial G(r,\phi,\tau)}{\partial\phi}\right|_{\phi=0}
-\left.\frac{\partial G(r,\phi,\tau)}{\partial\phi}\right|_{\phi=\alpha}\right).
\eeq
For a microscopic starting point,~(\ref{gsimple}) yields
\beq
f_{\tau,r}(\tau,r)\approx\frac{4\theta}{\pi\Gamma(2\theta)}
\left(\frac{a}{4D}\right)^{2\theta}\,\frac{r^{2\theta-1}}{\tau^{2\theta+1}}\,
\e^{-r^2/(4D\tau)}.
\label{ftaur}
\eeq
The expression~(\ref{ftau}) of the tail of the distribution of $\tau$
can be recovered by integrating the above result over $r$.
Similarly, integrating~(\ref{ftaur}) over $\tau$ yields the following expression
\beq
f_r(r)\approx\frac{4\theta}{\pi}\,\frac{a^{2\theta}}{r^{2\theta+1}}
\eeq
for the tail of the distribution of $r$.

For a given hitting time $\tau$,
the conditional distribution of the hitting distance $r$ reads
\beq
f_r(r|\tau)=\frac{f_{\tau,r}(\tau,r)}{f_\tau(\tau)},
\eeq
i.e.,
\beq
f_r(r|\tau)\approx\frac{2}{\Gamma(\theta)(4D\tau)^\theta}\,r^{2\theta-1}\,\e^{-r^2/(4D\tau)}.
\eeq
This result does not depend on the microscopic scale $a$ anymore.
Equivalently, setting
\beq
r=(4D\tau)^{1/2}\xi,
\label{rxi}
\eeq
in agreement with diffusive scaling,
the reduced variable $\xi$ has the universal distribution
\beq
f_\xi(\xi)=\frac{2}{\Gamma(\theta)}\,\xi^{2\theta-1}\,\e^{-\xi^2},
\label{fxires}
\eeq
depending only on the exponent $\theta$, i.e., on the wedge angle $\alpha$.

\subsection{Number of diagonal records}
\label{diagnumber}

This section is devoted to the number $N_t^{({\rm D})}$ of diagonal records at time $t$,
denoted as~$N_t$ for short throughout Section~\ref{diag}.
We successively investigate the mean value and the statistics of $N_t$.
The main emphasis will be on asymptotic results at large times.

Let us consider first the Polya walk.
Every realization of the walk generates an infinite
sequence of hitting times $\tau_1,\tau_2,\dots$
Hitting times are iid even integer random variables with the distribution
\beq
p(\tau)=\sum_{x=1}^\infty p(\tau,x).
\eeq
Hereafter we adopt the line of thought and the notations
of our earlier work on renewal processes~\cite{us}.
For a given time $t$,
the number of diagonal records is the unique integer~$N_t$ such that
$T_{(N_t)}\le t<T_{(N_t+1)}$, with the definition~(\ref{partials}).
This number is random, as it depends on the draw of the whole process $\{\tau_n\}$.
Let
\beq
p_n(t)=\prob{N_t=n}=\prob{T_{(n)}\le t<T_{(n+1)}}
\eeq
denote the probability that $N_t$ equals some integer $n$.
In particular, the probability of having no record up to time $t$
is nothing but the survival probability
\beq
p_0(t)=\prob{\tau_1>t}=S(t)=\sum_{\tau=t+1}^\infty p(\tau),
\label{P0exact}
\eeq
i.e., the probability that the walker has not yet hit the target at time $t$.

At large times,
it is legitimate to view $\tau$ as a continuous variable,
and to approximate the exact discrete distribution $p(\tau)$ by a continuous one with density $f_\tau(\tau)$.
Within this setting, namely renewal processes in continuous time,
many quantities can be determined explicitly in Laplace space.
Examples are given in~\ref{apprenew}.
The first quantity of interest is the distribution $p_n(t)$ of the number of records.
We have
\beqa
\h p_n(s)&=&\lap{t} p_n(t)
\nonumber\\
&=&
\int_0^\infty p_n(t)\e^{-st}\,\dd t
\nonumber\\
&=&
\left\langle\int_{T_{(n)}}^{T_{(n+1)}}\e^{-st}\,\dd t\right\rangle
\nonumber\\
&=&\left\langle\frac{1-\e^{-s\tau_{n+1}}}{s}\,\e^{-sT_{(n)}}\right\rangle
\nonumber\\
&=&\frac{1-\h f_\tau(s)}{s}\,\h f_\tau(s)^n
\label{nlap}
\eeqa
for all $n\ge0$.
We have in particular
\beq
\h p_0(s)=\frac{1-\h f_\tau(s)}{s},
\eeq
yielding
\beq
p_0(t)=S(t)=\int_t^\infty f_\tau(\tau)\dd\tau.
\eeq
This result coincides with~(\ref{sftau})
and is the continuum analogue of~(\ref{P0exact}).
The Laplace transform of the mean number of records at time $t$ reads
\beq
\lap{t}\mean{N_t}
=\sum_{n=0}^\infty n\h p_n(s)=\frac{\h f_\tau(s)}{s(1-\h f_\tau(s))}.
\label{navelap}
\eeq

From now on, we focus our attention onto distributions $f_\tau(\tau)$
with a power-law tail of the form~(\ref{ftau}),
restricting the exponent $\theta$ to the range $0<\theta<1$.
Diagonal and simultaneous records,
studied in Sections~\ref{diag} and~\ref{simul},
respectively correspond to $\theta=1/4$ and $\theta=1/3$.
We have then
\beq
p_0(t)\approx\frac{c}{\theta\,t^\theta}
\label{p0asy}
\eeq
and
\beq
\h f_\tau(s)\approx1-\frac{\Gamma(1-\theta)}{\theta}\,c\,s^\theta.
\label{lapf}
\eeq
For $0<\theta<1$, the mean hitting time $\mean{\tau}$ is divergent,
so that the renewal process does not equilibrate,
but rather keeps a sensitive memory of its initial state.
Many quantities exhibit large fluctuations,
some of them being scale invariant~\cite{us}.

\subsubsection{Mean number of diagonal records.}
\label{diagave}

The growth law of the mean number of records
for an arbitrary exponent $\theta<1$
can be derived by inserting~(\ref{lapf}) into~(\ref{navelap}),
and inverting the Laplace transform.
To leading order, we obtain the power-law growth
\beq
\mean{N_t}\approx A\,t^\theta,\quad
A=\frac{\sin\pi\theta}{\pi c}.
\label{ntheo}
\eeq

The problem at hand, namely diagonal records of the Polya walk,
maps onto the continuum theory of Section~\ref{brown}
for a wedge angle $\alpha=2\pi$ (cut plane), so that the survival exponent reads
\beq
\theta=\frac{1}{4}.
\eeq
The results~(\ref{ntheo}) therefore read
\beq
\mean{N_t}\approx A\,t^{1/4},\quad
A=\frac{1}{\pi\sqrt2\,c}.
\label{ntheodiag}
\eeq
This fourth-root law was announced in~(\ref{123}).

\begin{figure}[!ht]
\begin{center}
\includegraphics[angle=0,width=.6\linewidth,clip=true]{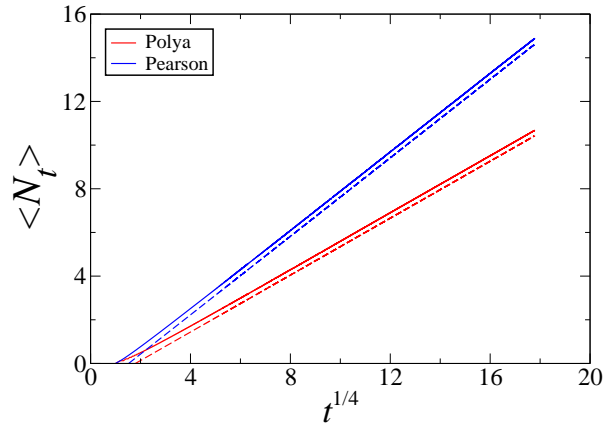}
\caption{\small
Full curves:
numerical data for mean number $\mean{N_t}$ of diagonal records
for Polya and Pearson walks (see legend) against $t^{1/4}$ up to $t=10^5$.
Dashed lines (slightly displaced for a better readability):
least-square fits of data for $t>10^3$.}
\label{numdiag}
\end{center}
\end{figure}

Figure~\ref{numdiag} shows numerical data
for the mean number $\mean{N_t}$ of diagonal records both for Polya and for Pearson walks
against $t^{1/4}$ up to $t=10^5$.
Both datasets exhibit a very accurate linear growth as a function of $t^{1/4}$.
The slopes $A$ of the least-square fits shown as dashed lines,
and the corresponding values of the tail parameter $c$ according to~(\ref{ntheodiag}),
are given in Table~\ref{acdiag} for both kinds of walks.

\begin{table}[!ht]
\begin{center}
\begin{tabular}{|l|c|c|c|}
\hline
walk & $A$ & $c$ & $c_0$ \cr
\hline
Polya & 0.652 & 0.345 & -- \cr
Pearson & 0.899 & 0.250 & 0.290 \cr
\hline
\end{tabular}
\caption
{Numerical values of the amplitude $A$ of the power-law growth~(\ref{ntheodiag})
of the mean number of diagonal records,
as extracted from the data shown in Figure~\ref{numdiag},
and of the corresponding tail parameter $c$ for Polya and Pearson walks.
For Pearson walks, the tail parameter $c_0$ of the first hitting time
(see Section~\ref{diagfull}) is also given.}
\label{acdiag}
\end{center}
\end{table}

For Polya walks,
diagonal records admit the recursive construction described
in Section~\ref{diagrecur}, and therefore exactly correspond to a renewal process.
It was therefore no surprise that the power law~(\ref{ntheodiag}) holds at large times.
For Pearson walks,
diagonal records do not admit such a construction stricto sensu.
The basic reason is that the starting point of the walk,
namely the origin, lies exactly on the diagonal,
whereas the steps that give rise to subsequent records,
i.e., those which cross the diagonal,
have their endpoints in a close vicinity of the diagonal, but not right onto it.
It was therefore not fully granted
that the power law~(\ref{ntheodiag}) would hold equally accurately for Pearson walks.
This growth law can however be expected on intuitive grounds to be universal,
i.e., to hold for all kinds of random walks in the diffusive universality class.
The amplitude $A$ is however not universal,
but rather depends on microscopic details of the walk.
Even in the diffusive theory describing a Brownian particle,
the tail parameter depends on the microscopic initial distance $a$ (see~(\ref{cbrown})),
which is a proxy for an effective lattice spacing or any other short-distance cutoff
in a walk consisting of discrete steps.

\subsubsection{Full statistics of number of diagonal records.}
\label{diagfull}

At large times,
the statistics of the number $N_t$ of records can be derived
by means of a scaling analysis of the exact expression~(\ref{nlap}).
Omitting details, we are left with the scaling formula~\cite{us}
\beq
N_t\approx\frac{\theta}{\Gamma(1-\theta)c}\,t^\theta X,
\label{nx}
\eeq
where the distribution of the reduced variable $X$ reads
\beq
f_X(X)=\int\frac{\dd z}{2\pi\ii}\,z^{\theta-1}\,\e^{z-Xz^\theta}.
\label{fcontour}
\eeq
We have the identity
\beq
X\equiv(L_\theta)^{-\theta},
\eeq
where $L_\theta$ is distributed according to the one-sided L\'evy stable law of index $\theta$
and a suitably chosen scale factor.
The density at $X=0$ and the mean value of $X$ read
\beqa
&&f_X(0)=\frac{1}{\Gamma(1-\theta)},
\label{forigin}
\\
&&\mean{X}=\frac{1}{\Gamma(1+\theta)}.
\eeqa
These results can be shown to be respectively in agreement with~(\ref{p0asy}) and~(\ref{ntheo}).

Whenever $\theta<1/2$,
the distribution of $X$ takes its maximum at $X=0$,
decays monotonically as a function of $X$,
and falls off as a stretched exponential,
\beq
f_X(X)\sim\exp\left(-(1-\theta)(\theta^\theta X)^{1/(1-\theta)}\right),
\eeq
so that all moments of $X$ are convergent.

For $\theta=1/4$, the distribution $f_X(X)$
is given by a linear combination of three hypergeometric functions
of type ${\null}_0F_2$~\cite{SM,Bar,Pen}.
The integral expression
\beq
f_X(X)=\frac{4}{\pi}\int_0^\infty\e^{-4y^4-Xy}(\cos Xy-\sin Xy)\dd y,
\label{fquart}
\eeq
obtained by folding the contour in~(\ref{fcontour}) onto the negative real axis,
setting $z=-4y^4$,
is more suitable for a numerical evaluation.

\begin{figure}[!ht]
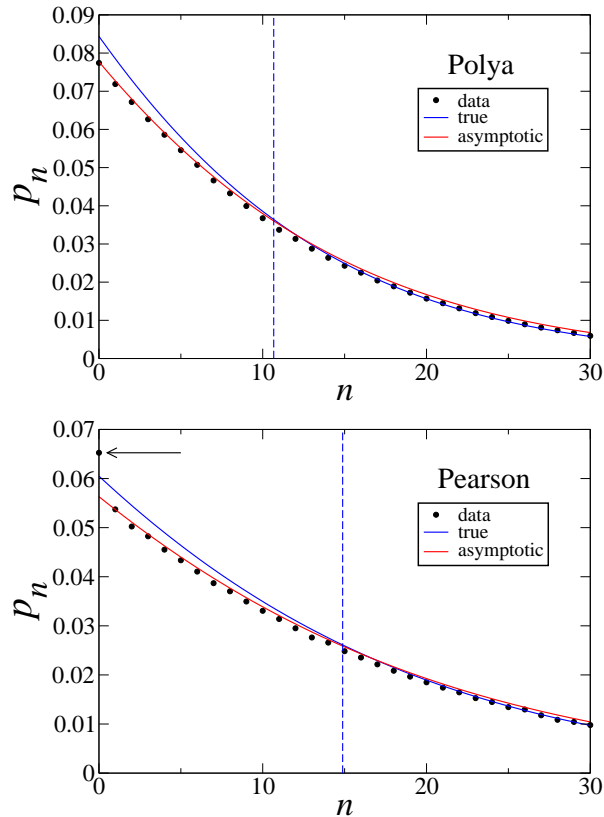

\begin{center}
\includegraphics[angle=0,width=.6\linewidth,clip=true]{hpolya.eps}
\vskip 6pt
\includegraphics[angle=0,width=.6\linewidth,clip=true]{hpearson.eps}
\caption{\small
Full distribution of number of diagonal records, $p_n(t)=\prob{N_t=n}$, at time~$t=10^5$
for Polya (top) and Pearson (bottom) walks.
Symbols: numerical data.
Blue (red) curves: true (asymptotic) theoretical predictions (see text).
Vertical dashed lines: mean values $\mean{N_t}\approx10.67$ (Polya)
and $\mean{N_t}\approx14.88$ (Pearson).}
\label{hdiag}
\end{center}
\end{figure}

Figure~\ref{hdiag} shows numerical data
for the distribution of the number of diagonal records, $p_n(t)=\prob{N_t=n}$,
of Polya and Pearson walks for $t=10^5$.
The mean record numbers,
$\mean{N_t}\approx10.67$ (Polya) and $\mean{N_t}\approx14.88$ (Pearson), are not very large,
so that sizeable corrections to scaling might be expected and are indeed observed.
In order to make a quantitative comparison between the plotted data and
the theoretical prediction~(\ref{nx}),~(\ref{fquart}),
the constant of proportionality between $n$ and $X$ has been determined in two ways.
Blue curves, labelled \textit{true}, are obtained
by using the true finite-time mean values given above and shown as vertical dashed lines.
Red curves, labelled \textit{asymptotic}, are obtained
by using the asymptotic growth law~(\ref{ntheodiag}),
with amplitudes $A$ given in Table~\ref{acdiag},
resulting in the estimates
$\mean{N_t}\approx11.59$ (Polya) and $\mean{N_t}\approx15.99$ (Pearson),
some 8 percent above true values.
It is observed that the distribution $p_n$ is better represented by the red curves
for small values of the record number $n$, and by the blue curves for large $n$.
A rather sharp crossover between both regimes is observed for values of $n$
comparable to the mean record number (vertical dashed lines).

Another striking observation can be made on the lower panel of Figure~\ref{hdiag}.
For Pearson walks, the probability $p_0$ (arrow) is significantly larger
than the extrapolation of the other data points.
This is again due to the fact that records of Pearson walks
are not given by a renewal process stricto sensu.
As a consequence, the survival probability starting from the origin falls off as
\beq
p_0(t)\approx\frac{c_0}{\theta\,t^\theta},
\label{p00}
\eeq
where the tail parameter $c_0$
of the distribution of the first hitting time of the diagonal
is a priori different from the tail parameter $c$ pertaining to the $n$th hitting time
for~$n$ large enough,
and entering the asymptotic results derived above.
Inserting the measured value $p_0\approx0.0652$ into~(\ref{p00}),
we obtain $c_0\approx0.290$.
This number, listed in Table~\ref{acdiag}, is some 16 percent higher than $c$.

\begin{figure}[!ht]
\begin{center}
\includegraphics[angle=0,width=.6\linewidth,clip=true]{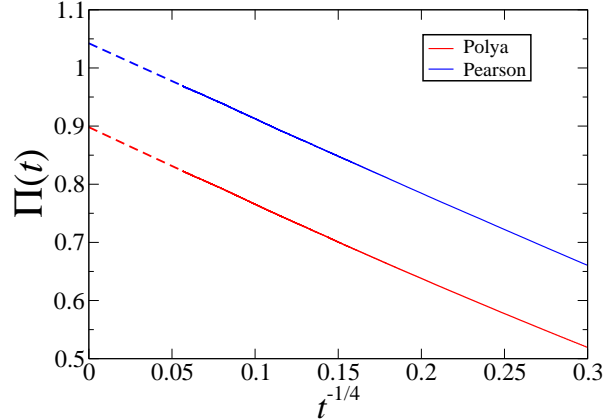}
\caption{\small
Full curves: numerical data for product $\Pi(t)$ (see~(\ref{pidef}))
for Polya and Pearson walks
(see legend) against $t^{-1/4}$.
Dashed curves: quadratic extrapolations (see text).}
\label{proddiag}
\end{center}
\end{figure}

Let us now take another perspective and consider the product
\beq
\Pi(t)=\mean{N_t}p_0(t)
\label{pidef}
\eeq
of the mean number of records at time $t$ by the probability
of having no record up to time $t$.
Renewal theory (see~(\ref{p0asy}),~(\ref{ntheo}))
predicts that this quantity converges to the limit
\beq
\Pi=\frac{Ac}{\theta}=\frac{\sin\pi\theta}{\pi\theta},
\eeq
i.e., $\Pi=2\sqrt{2}/\pi\approx0.900316$ for $\theta=1/4$.
Figure~\ref{proddiag} shows the product $\Pi(t)$
against~$t^{-1/4}$ for Polya and Pearson walks.
The rightmost data points correspond to $t=125$.
Both datasets are roughly parallel to each other,
and vary over an appreciable range,
confirming thus the importance of corrections to scaling.
Quadratic extrapolations (dashed curves)
yield the asymptotic values $\Pi\approx0.899$ for Polya walks,
in excellent agreement with the theoretical value stemming from renewal theory,
and $\Pi\approx1.042$ for Pearson walks.
Equating this number to $Ac_0/\theta$, with $A\approx0.899$ (see Table~\ref{acdiag}),
we consistently recover the result $c_0\approx0.290$ listed in Table~\ref{acdiag}.

\subsection{Epoch and location of current diagonal record}
\label{diagel}

In this section we investigate properties of the current (i.e., latest to date)
diagonal record at a fixed large time $t$.
Our goal is to derive the joint distribution of the epoch~$T_t$
and of the location of the current record, measured by its abscissa $\bar X_t$.
We have introduced this notation in order to distinguish
the current abscissa $X_t$ of the walker at time $t$ (see~(\ref{eq:Rt}))
and the abscissa $\bar X_t$ of the current record at time $t$,
i.e., equivalently,
the abscissa of the rightmost intersection of the first $t$ steps of the walk and of the diagonal.

Here again, the emphasis will be on asymptotic results at large times.
We consider first the Polya walk, for which the problem of records
exactly amounts to a renewal process.
For a given walk, we have (see~(\ref{partials}))
\beqa
T_t&=&T_{(N_t)}=\tau_1+\cdots+\tau_{N_t},
\nonumber\\
\bar X_t&=&X_{(N_t)}=x_1+\cdots+x_{N_t},
\label{txun}
\eeqa
where the number of terms in each sum is the number $N_t$ of records at time~$t$.

For a large time $t$,
the typical temporal increment scales as $\tau\sim t/\mean{N_t}\sim t^{3/4}$,
and the typical spatial increment scales as $x\sim\tau^{1/2}\sim t^{3/8}$.
All increments~$\tau_n$ and $x_n$ are therefore typically large,
so that we can have recourse to the continuum theory,
which predicts (see~(\ref{rxi}))
\beq
x_n\approx\sqrt{\frac{\tau_n}{2}}\;\xi_n,
\label{xxi}
\eeq
where the $\xi_n$ are independent from the $\tau_n$
and drawn from the distribution~(\ref{fxires}), with $\theta=1/4$.
In deriving~(\ref{xxi}), we have used $D=1/4$ (see~(\ref{d14})),
and the fact that the abscissa of a point along the diagonal is $x=r/\sqrt2$.

Diffusive scaling implies
\beq
T_t\approx t\,W,\quad
\bar X_t\approx\sqrt{\frac{t}{2}}\;U,
\label{wudef}
\eeq
where the dimensionless reduced variables $W$ and $U$ are distributed
according to some non-trivial joint distribution $f_{W,U}(W,U)$.
This distribution is expected to be universal
among all kinds of random walks in the diffusive universality class.
The variable $U$ has been normalized in order to avoid most factors of $\sqrt2$
in subsequent developments.

In order to determine the joint distribution of $W$ and $U$,
let us introduce the bivariate characteristic function
\beq
B(\sigma,p,t)=\bigmean{\e^{-\sigma T_t-p\sqrt2\,\bar X_t}}.
\eeq
Using~(\ref{txun}) and~(\ref{xxi}),
the above definition can be recast as
\beq
B(\sigma,p,t)=\bigmean{\prod_{n=1}^{N_t}b(\sigma,p,\tau_n)},
\eeq
with
\beq
b(\sigma,p,\tau)
=\bigmean{\e^{-\sigma\tau-p\sqrt{\tau}\xi}}
=\e^{-\sigma\tau}\int_0^\infty f_\xi(\xi)\e^{-p\sqrt\tau\xi}\,\dd\xi.
\label{bun}
\eeq
The quantity $B(\sigma,p,t)$
is therefore a multiplicative observable of the form~(\ref{bdef}), investigated in~\ref{apprenew}.
The Laplace transform of $B(\sigma,p,t)$
with respect to $t$ is therefore given by~(\ref{bres}), i.e.,
\beq
\h B(\sigma,p,s)=\frac{1-\h f_\tau(s)}{s(1-h_B(p,s+\sigma))}.
\label{bfullres}
\eeq
Let us estimate the above expression in the scaling regime where $p$, $s$ and $\sigma$ are small,
for an arbitrary exponent in the range $\theta<1/2$.
The numerator of~(\ref{bfullres}) is given by
\beq
1-\h f_\tau(s)\approx\frac{\Gamma(1-\theta)}{\theta}\,c\,s^\theta
\label{laps}
\eeq
(see~(\ref{lapf})), whereas the denominator involves the quantity
\beqa
1-h_B(p,s)
&\approx& 1-\int_0^\infty f_\tau(\tau)\e^{-s\tau}\,\dd\tau
\int_0^\infty f_\xi(\xi)\e^{-p\sqrt\tau\xi\,}\dd\xi
\nonumber\\
&\approx&-\int_0^\infty S(\tau)\dd\tau\,
\frac{\dd}{\dd\tau}\left(\e^{-s\tau}
\!\int_0^\infty f_\xi(\xi)\e^{-p\sqrt\tau\xi}\,\dd\xi\right).
\label{bdeno}
\eeqa
The second expression, obtained by means of an integration by parts,
is suitable for an explicit evaluation in the scaling regime.
This is performed in~\ref{appderiv} and yields (see~(\ref{appbsca}))
\beq
\h B(\sigma,p,s)\approx\frac{(1+y)^{-\theta}}{s\,\phi((1+y)^{-1/2}z)},
\label{bsca}
\eeq
where the scaling variables $y$ and $z$ read
\beq
y=\frac{\sigma}{s},\quad z=\frac{p}{\sqrt{s}},
\eeq
and the scaling function $\phi(\zeta)$ is obtained
in parametric form as (see~(\ref{appphi}))
\beq
\phi=\frac{\cos 2\theta\gamma}{\cos\pi\theta},\quad
\zeta=2\cos\gamma.
\eeq
This closes our analysis for arbitrary values of $\theta$.

In the present situation ($\theta=1/4$),
we have the explicit expression (see~(\ref{appphi4}))
\beq
\phi(\zeta)=\left(1+\frac{\zeta}{2}\right)^{1/2}.
\label{phi4}
\eeq
Inserting this form into~(\ref{bsca}), we obtain
\beq
\h B(\sigma,p,s)\approx\sqrt{2}\,s^{-3/4}\left(p+2\sqrt{s+\sigma}\right)^{-1/2}.
\label{bfull}
\eeq
The triple inverse Laplace transform of the above expression
can be worked out explicitly by elementary means.
Inverting successively over $p$, $\sigma$ and $s$,
we obtain the joint distribution of $W$ and $U$ in the form
\beq
f_{W,U}(W,U)=\frac{\Gamma(1/4)}{\pi^2}\,\frac{\sqrt{U}\,\e^{-U^2/W}}{W^{3/2}(1-W)^{1/4}}.
\label{fwu}
\eeq

The distributions of $W$ and of $U$ can be derived by integrating~(\ref{fwu}) over the other variable.

The distribution of $W$ reads
\beq
f_W(W)=\frac{1}{\pi\sqrt2}\,W^{-3/4}(1-W)^{-1/4}\quad(0<W<1).
\eeq
We have thus recovered (for $\theta=1/4$)
the beta distribution of the reduced epoch $W=T_t/t$ of the last renewal
for arbitrary $\theta<1$ (see e.g.~\cite{us}), i.e.,
\beq
f_W(W)=\frac{\sin\pi\theta}{\pi}\,W^{-(1-\theta)}(1-W)^{-\theta}\quad(0<W<1).
\label{wres}
\eeq

The distribution of $U$ reads
\beq
f_U(U)=\frac{\sqrt2}{\pi^{3/2}}\,\e^{-U^2/2}\,K_{1/4}(U^2/2),
\label{fures}
\eeq
where $K_{1/4}$ is the modified Bessel function.
The behavior of this distribution at small and large values of $U$ reads
\beqa
f_U(U)&\approx&\frac{\Gamma(1/4)}{\pi^{3/2}\,\sqrt{U}}\quad(U\to0),
\nonumber\\
f_U(U)&\approx&\frac{\sqrt2}{\pi U}\,\e^{-U^2}\quad(U\to\infty).
\eeqa

All joint moments of $W$ and $U$ can also be derived from~(\ref{fwu}).
They read
\beq
\mean{W^mU^n}=\frac{1}{\pi\sqrt2}\,\frac{\Gamma(m+n/2+1/4)\Gamma(n/2+3/4)}{\Gamma(m+n/2+1)}.
\eeq
We have in particular
\beqa
\mean{W}=\frac{1}{4},\quad
\mean{W^2}=\frac{5}{32},\quad
\mean{W^3}=\frac{15}{128},\quad
\mean{W^4}=\frac{195}{2048},
\nonumber\\
\mean{U}=\frac{1}{2\sqrt\pi},\quad
\mean{U^2}=\frac{3}{16},\quad
\mean{U^3}=\frac{5}{16\sqrt\pi},\quad
\mean{U^4}=\frac{105}{512},
\nonumber\\
\mean{WU}=\frac{1}{4\sqrt\pi},\quad
\mean{W^2U}=\frac{7}{40\sqrt\pi},\quad
\mean{WU^2}=\frac{15}{128}.
\eeqa

The above value of $\mean{U}$ implies
\beq
\mean{\bar X_t}\approx\sqrt{\frac{t}{8\pi}}.
\label{xunave}
\eeq
This result is $\sqrt{8}$ times smaller than the mean absolute abscissa
of the walker at time~$t$, $\mean{\abs{X_t}}\approx\sqrt{t/\pi}$.

The reduced epoch $W$ and abscissa $U$ of the current record
are significantly correlated.
Their correlation coefficient indeed reads
\beq
C=\frac{\mean{WU}_c}{\left(\mean{W^2}_c\mean{U^2}_c\right)^{1/2}}
=\left(\frac{8}{3(3\pi-4)}\right)^{1/2}\approx0.701121.
\eeq

\begin{figure}[!ht]
\begin{center}
\includegraphics[angle=0,width=.6\linewidth,clip=true]{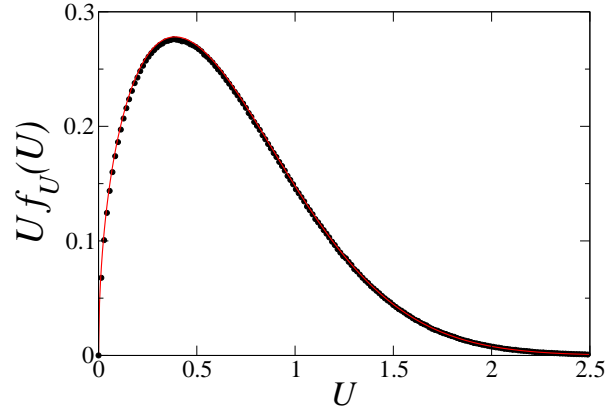}
\caption{\small
Black symbols: Distribution of abscissa $\bar X_t$ of current diagonal record
of Polya walks at time $t=10^4$, rescaled according to~(\ref{wudef}).
Full red curve: asymptotic prediction $f_U(U)$ (see~(\ref{fures})).
Both quantities are multiplied by $U$ (see text).}
\label{hx}
\end{center}
\end{figure}

Figure~\ref{hx} shows a comparison between the distribution of the (integer) abscissa~$\bar X_t$
of the current record of Polya walks at time $t=10^4$,
rescaled according to~(\ref{wudef}), and the asymptotic prediction~(\ref{fures}).
The product~$Uf_U(U)$ is plotted in order to better reveal the features of the distribution.
A very good agreement is obtained without any adjustable parameter.
Corrections to scaling are very small (of the order of one percent).
This situation is in strong contrast with
the statistics of the number of records, displayed in Figure~\ref{hdiag}.

\section{Simultaneous records}
\label{simul}

\subsection{Recursive construction}
\label{simulrecur}

\begin{figure}[!ht]
\begin{center}
\includegraphics[angle=0,width=.4\linewidth,clip=true]{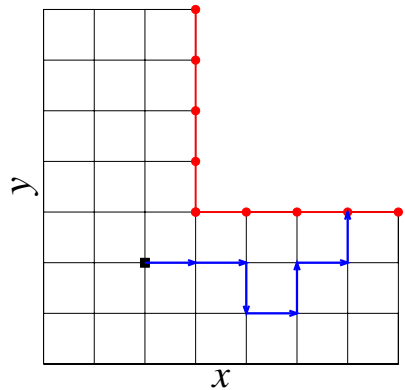}
\caption{\small
First step of recursive construction of simultaneous records of a Polya walk.
Black square: origin.
Red: target (quadrant).
Blue: Polya walk.}
\label{cstrsimul}
\end{center}
\end{figure}

This section is devoted to the statistics of simultaneous records,
shown as blue symbols in Figure~\ref{exemples}.
Let us begin by considering the Polya walk.
Simultaneous records are germane to diagonal ones, investigated in Section~\ref{diag},
in the sense that they also admit a recursive description,
whose first step is illustrated in Figure~\ref{cstrsimul}.
The target is the quadrant issued from the point $(1,1)$, marked in red.
The first simultaneous record corresponds to the first hitting of the target
by a walk issued from the origin.
On the example, the walk makes $\tau_1=7$ steps
before it hits the target at the point $(x_1=4,y_1=1)$.

The second record can be constructed by considering
the location $(x_1,y_1)=(4,1)$ of the first record as a new origin.
The walk issued from that origin makes $\tau_2$ steps
before it hits the quadrant at some point $(x_2,y_2)$, and so on.
We have therefore reduced the problem to a renewal process.
The $n$th simultaneous record takes place at time $T_{(n)}$
and at the lattice point $\vec R_{(n)}=(X_{(n)},Y_{(n)})$, where
\beqa
T_{(n)}&=&\tau_1+\cdots+\tau_n,
\nonumber\\
X_{(n)}&=&x_1+\cdots+x_n,
\nonumber\\
Y_{(n)}&=&y_1+\cdots+y_n.
\label{simulpartials}
\eeqa

Temporal and spatial increments $(\tau_n,x_n,y_n)$
are iid triples of integer random variables.
Their joint distribution $p(\tau,x,y)$ identifies with the distribution
of the hitting time~$\tau$ and of the coordinates $(x,y)$ of the hitting point of the quadrant
by a random walk starting at the origin, as shown in Figure~\ref{cstrsimul}.
If the walker hits the horizontal part of the boundary of the quadrant,
as illustrated in Figure~\ref{cstrsimul},
$x\ge1$ can be an arbitrary integer and $y=1$.
If the walker hits the vertical part of the boundary,
$x=1$ and $y\ge1$ can be an arbitrary integer.
If the walker hits the tip of the boundary, we have $x=y=1$.
Finally, the hitting time $\tau$ has the same parity as the sum $x+y$.

Throughout the following,
it will be sufficient to know the asymptotic behavior
of the joint distribution $p(\tau,x,y)$ when all variables are large.
This asymptotic form can again be derived
by means of the continuum diffusion theory of Section~\ref{brown}.
The present problem maps onto the survival of a Brownian particle
in a wedge with angle $\alpha=3\pi/2$ (complement of a quadrant),
so that the survival exponent reads
\beq
\theta=\frac{1}{3}.
\eeq

\subsection{Mean number of simultaneous records}
\label{simulave}

This section is devoted to the number $N_t^{({\rm S})}$ of simultaneous records at time $t$,
denoted as $N_t$ for short throughout Section~\ref{simul}.
We focus our attention onto the mean number of records.
The asymptotic growth of this quantity is given
by the prediction~(\ref{ntheo}) of renewal theory with $\theta=1/3$,
i.e.,
\beq
\mean{N_t}\approx A\,t^{1/3},\quad
A=\frac{\sqrt3}{2\pi c}.
\label{ntheosimul}
\eeq
This third-root law was announced in~(\ref{123}).

\begin{figure}[!ht]
\begin{center}
\includegraphics[angle=0,width=.6\linewidth,clip=true]{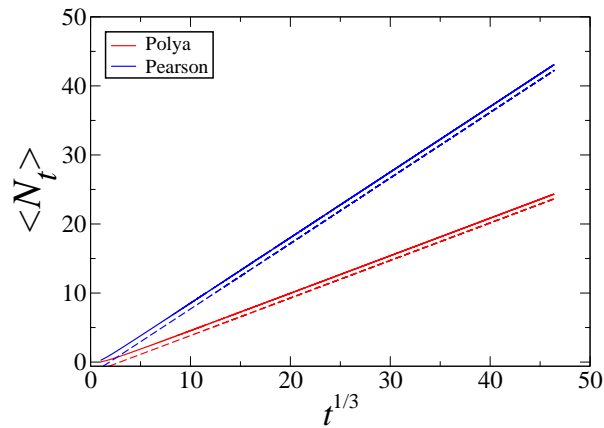}
\caption{\small
Full curves:
numerical data for mean number $\mean{N_t}$ of simultaneous records
for Polya and Pearson walks (see legend) against $t^{1/3}$ up to $t=10^5$.
Dashed lines (slightly displaced for a better readability):
least-square fits of data for $t>10^3$.}
\label{numsimul}
\end{center}
\end{figure}

Figure~\ref{numsimul} shows numerical data
for the mean number $\mean{N_t}$ of simultaneous records both for Polya and for Pearson walks
against $t^{1/3}$ up to $t=10^5$.
Both datasets exhibit a very accurate linear behavior.
The slopes $A$ of the least-square fits shown as dashed lines,
and the corresponding values of the tail parameter $c$ according to~(\ref{ntheosimul}),
are given in Table~\ref{acsimul} for both kinds of walks.

\begin{table}[!ht]
\begin{center}
\begin{tabular}{|l|c|c|}
\hline
walk & $A$ & $c$ \cr
\hline
Polya & 0.544 & 0.507 \cr
Pearson & 0.949 & 0.290 \cr
\hline
\end{tabular}
\caption
{Numerical values of amplitude $A$ of the power-law growth~(\ref{ntheosimul})
of mean number of simultaneous records,
as extracted from the data shown in Figure~\ref{numsimul},
and of corresponding tail parameter $c$ for Polya and Pearson walks.}
\label{acsimul}
\end{center}
\end{table}

At variance with the situation of diagonal records (see Section~\ref{diagfull}),
data on the full distribution of the number of simultaneous records for Pearson walks (not shown)
do not hint at any measurable difference between
the tail parameter $c_0$ of the distribution of the first hitting time (see~(\ref{p00}))
and the tail parameter $c$ of late hitting times,
entering asymptotic results from renewal theory.

\subsection{Location of current simultaneous record}
\label{simulloc}

In this section we investigate properties of the location
$\bar{\vec R}_t=(\bar X_t,\bar Y_t)$ of the current
simultaneous record at a fixed large time $t$.
Here again, the emphasis will be on asymptotic results at large times,
and we consider first the Polya walk, for which the problem of records
exactly amounts to a renewal process.
For a given walk, we have (see~(\ref{simulpartials}))
\beqa
T_t&=&T_{(N_t)}=\tau_1+\cdots+\tau_{N_t},
\nonumber\\
\bar X_t&=&X_{(N_t)}=x_1+\cdots+x_{N_t},
\nonumber\\
\bar Y_t&=&Y_{(N_t)}=y_1+\cdots+y_{N_t}.
\label{txde}
\eeqa
The number of terms in each sum is the number $N_t$ of records at time~$t$.

At large times,
the typical temporal increment scales as $\tau\sim t/\mean{N_t}\sim t^{2/3}$,
and the typical spatial increment scales as $x\sim\tau^{1/2}\sim t^{1/3}$.
All increments are typically large,
so that the spatial increments $x_n$ and $y_n$ can be estimated by means of~(\ref{rxi}),
with $D=1/4$ (see~(\ref{d14})).
More precisely:
if the walker hits the horizontal part of the boundary of the quadrant,
we have $x_n\approx\sqrt{\tau_n}\,\xi_n$,
where $\xi_n$ is distributed according to~(\ref{fxires}),
with $\theta=1/3$,
whereas $y_n=1$ is negligible;
if the walker hits the vertical part of the boundary,
we have $y_n\approx\sqrt{\tau_n}\,\xi_n$,
whereas $x_n=1$ is negligible.
Both events are related to each other by symmetry.
In particular, they are equally probable.
The event where $x_n=y_n=1$ has negligible weight.
All in all, we have
\beq
\left\{
\matrix{
x_n\approx\sqrt{\tau_n}\,\xi_n, & y_n\approx0\hfill &\hbox{with prob.~1/2},\cr
x_n\approx0,\hfill & y_n\approx\sqrt{\tau_n}\,\xi_n\quad &\hbox{with prob.~1/2}.
}
\right.
\label{xyxi}
\eeq

Diffusive scaling implies the asymptotic forms
\beq
\bar X_t\approx U\,\sqrt{t},\quad
\bar Y_t\approx V\,\sqrt{t},
\label{uvdef}
\eeq
where the reduced variables $U$ and $V$ are distributed
according to some non-trivial symmetric joint distribution $f_{U,V}(U,V)$.
This distribution is again expected to be universal.
In order to investigate it, we introduce the characteristic function
\beq
B(p,q,t)=\bigmean{\e^{-p\bar X_t-q\bar Y_t}}.
\eeq
Using~(\ref{txde}) and~(\ref{xyxi}),
the above definition can be recast as
\beq
B(p,q,t)=\bigmean{\prod_{n=1}^{N_t}b(p,q,\tau_n)},
\eeq
with
\beqa
b(p,q,\tau)
&=&\frac12\bigmean{\e^{-p\sqrt\tau\xi}+\e^{-q\sqrt\tau\xi}}
\nonumber\\
&=&\frac12\int_0^\infty f_\xi(\xi)\left(\e^{-p\sqrt\tau\xi}+\e^{-q\sqrt\tau\xi}\right)\dd\xi.
\eeqa
The quantity $B(p,q,t)$
is therefore a multiplicative observable of the form~(\ref{bdef}), investigated in~\ref{apprenew}.
A comparison with~(\ref{bun}) (with $\sigma=0$) yields
\beq
\h B(p,q,s)\approx\frac{2}{s(\phi(y)+\phi(z))},\quad
y=\frac{p}{\sqrt{s}},\quad z=\frac{q}{\sqrt{s}}.
\label{bdesca}
\eeq

In the present situation ($\theta=1/3$),
the scaling function $\phi(\zeta)$ obeys the equation of a so-called unicursal cubic
(see~(\ref{appphi3})), i.e.,
\beq
\zeta^2=(\phi-1)^2(\phi+2).
\label{phi3}
\eeq
Its power-law expansion near $\zeta=0$ reads
\beq
\phi(\zeta)
=1+\frac{\zeta}{\sqrt{3}}-\frac{\zeta^2}{18}+\frac{5\zeta^3}{216\sqrt{3}}-\frac{\zeta^4}{243}
+\frac{77\zeta^5}{31104\sqrt{3}}-\frac{7\zeta^6}{13122}
+\cdots
\label{phiser}
\eeq

At variance with the case of diagonal records,
we have not been able to derive from~(\ref{bdesca})
a closed-form expression for the joint distribution $f_{U,V}(U,V)$.
The joint moments
\beq
\mu_{m,n}=\mean{U^mV^n}
\eeq
can however be investigated as follows.
The characteristic function $B(p,q,t)$ and its Laplace transform read
\beqa
B(p,q,t)\approx\sum_{m,n=0}^\infty\frac{(-p)^m}{m!}\,\frac{(-q)^n}{n!}\,t^{(m+n)/2}\,\mu_{m,n},
\nonumber\\
\h B(p,q,s)\approx\sum_{m,n=0}^\infty\frac{(-p)^m}{m!}\,\frac{(-q)^n}{n!}\,
\frac{\Gamma(1+(m+n)/2)}{s^{1+(m+n)/2}}\,\mu_{m,n}.
\eeqa
The expression~(\ref{bdesca}) therefore amounts to
\beq
\frac{2}{\phi(y)+\phi(z)}=\sum_{m,n=0}^\infty\frac{(-y)^m}{m!}\,
\frac{(-z)^n}{n!}\Gamma(1+(m+n)/2)\mu_{m,n}.
\label{uvmoms}
\eeq
Using~(\ref{phiser}) to expand the left-hand side of~(\ref{uvmoms}) as a bivariate power series,
we obtain the first few joint moments of $U$ and $V$:
\beqa
\mean{U}=\mean{V}=\frac{1}{\sqrt{3\pi}},
\nonumber\\
\mean{U^2}=\mean{V^2}=\frac{2}{9},\quad
\mean{UV}=\frac{1}{6},
\nonumber\\
\mean{U^3}=\mean{V^3}=\frac{35}{54\sqrt{3\pi}},\quad
\mean{U^2V}=\mean{UV^2}=\frac{11}{27\sqrt{3\pi}},\quad
\\
\mean{U^4}=\mean{V^4}=\frac{20}{81},\quad
\mean{U^3V}=\mean{UV^3}=\frac{59}{432},\quad
\mean{U^2V^2}=\frac{37}{324}.
\nonumber
\eeqa

The mean coordinates of the current simultaneous record,
$\mean{\bar X_t}=\mean{\bar Y_t}\approx\sqrt{t/(3\pi)}$,
are $\sqrt{3}$ times smaller than the mean absolute coordinates
of the walker at time~$t$, $\mean{\abs{X_t}}=\mean{\abs{Y_t}}\approx\sqrt{t/\pi}$.

The reduced coordinates $U$ and $V$ of simultaneous records
are significantly correlated.
Their correlation coefficient indeed reads
\beq
C=\frac{\mean{UV}_c}{\mean{U^2}_c}
=\frac{3(\pi-2)}{2(2\pi-3)}\approx0.521563.
\label{cres}
\eeq

\subsection{Angular distribution of current simultaneous record}
\label{simulang}

It can be expected on intuitive grounds
that simultaneous records tend to cluster near the diagonal,
rather than being uniformly distributed over the quadrant.
This picture is corroborated by the large value
of the correlation coefficient $C$ (see~(\ref{cres})).

In order to further elaborate in this direction,
we consider the angular distribution of simultaneous records,
i.e., the distribution of the polar angle~$\Phi_t$,
such that the ratio of the coordinates of the current simultaneous record at time $t$
is parametrized as
\beq
\lam_t=\frac{\bar Y_t}{\bar X_t}=\tan\Phi_t\quad(0<\Phi_t<\pi/2).
\eeq
The distribution of the ratio $\lam_t$ is studied in~\ref{appratio} and given by~(\ref{fratio}), i.e.,
\beq
f_{\lam_t}(\lam_t)=-\int\frac{\dd q}{2\pi\ii}\left.\frac{\partial B(p,q,t)}{\partial p}\right|_{p=-\lam_t q}.
\label{fratiode}
\eeq

In the scaling regime of large times, the expression~(\ref{bdesca})
demonstrates that $B(p,q,t)$ is asymptotically
a function of $P=p\sqrt{t}$ and $Q=q\sqrt{t}$,
so that~(\ref{fratiode}) becomes independent of $t$.
This is in agreement with the scaling law~(\ref{uvdef}), implying that $\lam_t$ approaches
\beq
\lam=\frac{V}{U}=\tan\Phi
\eeq
at large times.
The ratio $\lam$ and the polar angle $\Phi$ are therefore expected
to have non-trivial asymptotic distributions,
related to each other as
\beq
f_\Phi(\Phi)=\frac{f_\lam(\tan\Phi)}{\cos^2\Phi}.
\label{lamphi}
\eeq
The symmetry~(\ref{lamflam}), i.e.,
\beq
\lam f_\lam(\lam)=\frac{1}{\lam}\,f_\lam\!\left(\frac{1}{\lam}\right),
\eeq
translates to the expected symmetry
\beq
f_\Phi(\Phi)=f_\Phi(\pi/2-\Phi).
\eeq

From a quantitative viewpoint, using the scaling law~(\ref{bdesca}) valid at large times,
we can recast~(\ref{fratiode}) into the form
\beq
\lam f_\lam(\lam)=2\int\frac{\dd z}{2\pi\ii}\,\frac{\phi'(z)}{(\phi(z)+\phi(-\lam z))^2},
\label{fint}
\eeq
where the accent denotes a derivative.

For generic values of the ratio $\lam$,
setting $\theta=1/3$ and $\gamma=(\pi-3\ii\alpha)/2$ in~(\ref{appphi}),
with $\alpha$ real, we obtain the following hyperbolic parametrization
\beq
z=2\ii\sinh\frac{3\alpha}{2},\quad\phi=\cosh\alpha+\ii\sqrt3\sinh\alpha
\eeq
of the function $\phi(z)$ when $z$ runs over the imaginary axis.
This yields after some algebra the following integral representation of $f_\lam(\lam)$,
which is suitable for a numerical evaluation:
\beq
\lam f_\lam(\lam)=\frac{4\sqrt3}{\pi}\int_0^\infty\frac{N(\alpha)}{D(\alpha)^2}\,\dd\alpha,
\label{abint}
\eeq
with
\beqa
N(\alpha)&=&(4\cosh\alpha+\cosh\beta)(1+\sinh\alpha\sinh\beta)
\nonumber\\
&-&\cosh\alpha(2\cosh^2\alpha+\cosh^2\beta),
\nonumber\\
D(\alpha)&=&(\cosh\alpha+\cosh\beta)^2+3(\sinh\alpha-\sinh\beta)^2.
\eeqa
The dependence of the above result on the ratio $\lam$ is entirely encoded
in the definition of the implicit function $\beta(\lam,\alpha)$, such that
\beq
\sinh\frac{3\beta}{2}=\lam\sinh\frac{3\alpha}{2}.
\eeq

For $\lam=1$, i.e., $\Phi=\pi/4$, we have $\beta=\alpha$.
This is the only situation where the integral~(\ref{abint}) is elementary, yielding
\beq
f_\lam(1)=\frac{\sqrt3}{4},\quad
f_\Phi(\pi/4)=\frac{\sqrt3}{2}.
\eeq

In the regime where the ratio $\lam$ is large,
the most efficient route to derive the tail of $f_\lam(\lam)$
consists in coming back to~(\ref{fint}),
using the asymptotic form $\phi(-\lam z)\approx(-\lam z)^{2/3}$,
and changing variable from $z$ to $\phi$.
We thus obtain
\beqa
f_\lam(\lam)&\approx&\frac{2}{\lam^{7/3}}\int\frac{\dd\phi}{2\pi\ii}(\phi+2)^{-2/3}(1-\phi)^{-4/3}
\nonumber\\
&\approx&\frac{2\sqrt3}{\pi\lam^{7/3}}
\;\underbrace{\int_1^\infty(\phi+2)^{-5/3}(\phi-1)^{-1/3}\,\dd\phi}_{1/2},
\eeqa
where the second integral is derived from the first one
by performing an integration by parts and folding the contour onto the real axis.
We thus obtain the power-law estimates
\beqa
f_\lam(\lam)\approx\frac{\sqrt3}{\pi}\,\lam^{1/3}\quad(\lam\to0),
\nonumber\\
f_\lam(\lam)\approx\frac{\sqrt3}{\pi}\,\lam^{-7/3}\quad(\lam\to\infty),
\\
f_\Phi(\Phi)\approx\frac{\sqrt3}{\pi}\,\Phi^{1/3}\quad(\Phi\to0),
\nonumber\\
f_\Phi(\Phi)\approx\frac{\sqrt3}{\pi}\,(\pi/2-\Phi)^{1/3}\quad(\Phi\to\pi/2).
\eeqa

\begin{figure}[!ht]
\begin{center}
\includegraphics[angle=0,width=.6\linewidth,clip=true]{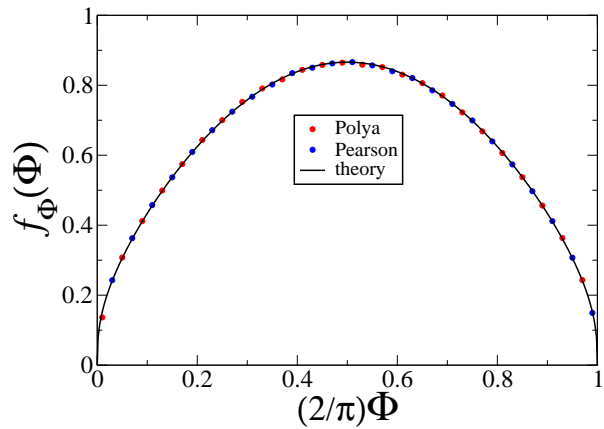}
\caption{\small
Symbols: histogram plots of distribution of polar angle~$\Phi$
of current simultaneous record of Polya and Pearson walks (see legend) at time $t=10^5$.
Full black curve: asymptotic theoretical prediction $f_\Phi(\Phi)$ (see~(\ref{lamphi}),~(\ref{abint})).}
\label{hangular}
\end{center}
\end{figure}

Figure~\ref{hangular} shows the distribution of the polar angle~$\Phi$
of the current record of Polya (red) and Pearson (blue) walks at time $t=10^5$.
Each dataset
contains 50 bins.
Every second bin of each dataset is plotted alternatively.
Both histograms are in very good agreement with the asymptotic theoretical prediction
$f_\Phi(\Phi)$ (see~(\ref{lamphi}),~(\ref{abint})), shown as a full curve.
Corrections to scaling are too small to be detected.

\section{Radial records}
\label{rad}

\subsection{General analysis}
\label{radgen}

This section is devoted to the statistics of radial records,
shown as green symbols in Figure~\ref{exemples}.
At variance with diagonal and simultaneous records,
investigated in Sections~\ref{diag} and~\ref{simul},
radial records do not admit any recursive construction
involving the hitting of translated copies of a target.
In other words, there is no underlying simple renewal process.

The following analysis is therefore partly heuristic.
Consider isotropic random walks consisting of discrete steps of unit length, in any dimension $d\ge2$.
Numerical simulations will be performed for Polya walks
(simple random walks on the hyper\-cubic lattice)
and Pearson walks (steps having unit length and uniformly random orientations).
Both kinds of walks obey $\mean{\vec R_t^2}=t$,
so that their diffusion coefficient reads
\beq
D=\frac{1}{2d}.
\label{diff}
\eeq
Let $\vec R_{(n)}$ be the positions of the walker at the successive radial records,
$R_{(n)}=\abs{\vec R_{(n)}}$ the corresponding radii, and $T_{(n)}$ the corresponding epochs.
In other words, $\vec R_{(n+1)}$ is the first position of the walker
which lies outside the sphere with radius $R_{(n)}$ centered at the origin.
For walks made of unit steps, we have
\beq
0<R_{(n+1)}-R_{(n)}\le1
\label{dr}
\eeq
in full generality, with formally $R_0=0$.
For both kinds of walks, we have $R_{(1)}=1$ and $T_{(1)}=1$.
For Polya walks we have either $R_{(2)}=2$ or $R_{(2)}=\sqrt2$,
whereas $T_{(2)}$ is an even integer, as the hypercubic lattice is bipartite in any dimension.
For Pearson walks, $R_{(2)}$ can already take any value between 1 and~2.

For a fixed time $t$, the number $N_t^{({\rm R})}$ of simultaneous records is denoted as $N_t$ for short
throughout Section~\ref{rad}.
The radius
\beq
\bar R_t=R_{(N_t)}
\eeq
of the current radial record
is nothing but the largest radius reached by the walk up to time $t$.

At large times,
the radius $R_t=\abs{\vec R_t}$ of the walk becomes the radius of a $d$-dimensional Brownian motion.
The latter process is known as a Bessel process of order $\nu=(d-2)/2$.
The largest radius $\bar R_t$ becomes the maximum of that process up to time $t$.
Diffusive scaling implies
\beq
R_t\approx S\sqrt{t},\quad
\bar R_t\approx U\sqrt{t}.
\label{sm}
\eeq
The reduced variables $S$ (associated with the radius of a generic point of the walk)
and $U$ (associated with the current maximal radius of the walk) have universal distributions,
which only depend on dimension $d$.
The distribution of $S$ is simply that of the radial part of
an isotropic $d$-dimensional Gaussian vector, normalized in accordance with~(\ref{diff}).
This reads
\beq
f_S(S)=\frac{2(d/2)^{d/2}}{\Gamma(d/2)}\,S^{d-1}\e^{-(d/2)S^2}.
\label{fs}
\eeq
We have in particular
\beq
\mean{S}=\sqrt{\frac{2}{d}}\frac{\Gamma((d+1)/2)}{\Gamma(d/2)},\quad\mean{S^2}=1.
\label{smoms}
\eeq
The distribution of the reduced variable $U$ is non-trivial.
It is known in the form of an infinite series involving the zeros
of the Bessel function~$J_0$ in two dimensions~\cite[p.~280]{borodin},
and more generally~$J_\nu$ in higher dimensions~\cite[p.~369]{borodin}.

The statistics of the number $N_t$ of radial records can be estimated as follows.
The difference between the radii of any two successive records
ought to average to some microscopic length (see~(\ref{dr}))
\beq
a=\lim_{n\to\infty}\bigmean{R_{(n+1)}-R_{(n)}}.
\eeq
It is therefore legitimate to expect that, for a given walk,
the number $N_t$ of records and the largest radius $\bar R_t$
are asymptotically proportional to each other, as
\beq
\bar R_t\approx a N_t.
\label{propor}
\eeq
We thus predict the scaling laws
\beq
N_t\approx\frac{U}{a}\,\sqrt{t},
\label{nfullrad}
\eeq
and in particular
\beq
\mean{N_t}\approx A\,\sqrt{t},
\label{nrad}
\eeq
with
\beq
A=\frac{\mean{U}}{a}.
\label{anrad}
\eeq
The square-root growth law~(\ref{nrad}) is well-known in the one-dimensional case (see~\cite{gmsrev}).
It was announced in~(\ref{123}) in the two-dimensional situation.
It is actually super-universal, in the sense that it holds in any spatial dimension.
This finding corroborates earlier numerical results in one, two and three dimensions~\cite{edery}.

To close, let us stress that
the distribution of the random variable $U$ and its mean value $\mean{U}$
entering the numerators of~(\ref{nfullrad}) and~(\ref{anrad}) are universal,
as they only depend on dimension $d$,
whereas the distance $a$ entering the denominators
depends a priori on the microscopic structure of the walk.

\subsection{Two dimensions}
\label{rad2}

We begin by illustrating the above general results in the two-dimensional case.
In this situation, the distribution of $U$ reads~\cite[p.~280]{borodin}
\beq
f_U(U)=\frac{1}{U^3}\sum_{k=1}^\infty\frac{j_k}{J_1(j_k)}\,\e^{-j_k^2/(4U^2)},
\label{fm}
\eeq
where $j_k$ are the zeros of the Bessel function $J_0$, growing as $j_k\approx(k-1/4)\pi$,
whereas $J_1(j_k)$ are the values of the Bessel function $J_1$ at these zeros.
We have in particular
\beqa
\mean{U}=\frac{1}{\sqrt\pi}\int_0^\infty\frac{\dd x}{I_0(x)}\approx1.175338,
\label{m1}
\\
\mean{U^2}=\frac{1}{2}\int_0^\infty\frac{x\,\dd x}{I_0(x)}\approx1.534414,
\eeqa
where $I_0$ is the modified Bessel function.
These numbers are to be compared with
$\mean{S}=\sqrt\pi/2\approx0.886226$ and $\mean{S^2}=1$ (see~(\ref{smoms})).

\begin{figure}[!ht]
\begin{center}
\includegraphics[angle=0,width=.6\linewidth,clip=true]{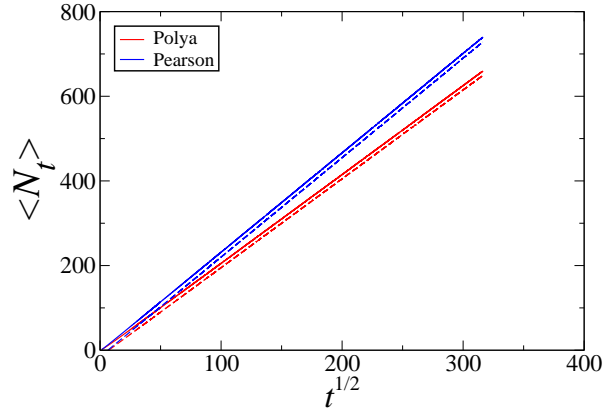}
\caption{\small
Full curves:
numerical data for mean number $\mean{N_t}$ of radial records
for Polya and Pearson planar walks (see legend) against $\sqrt{t}$ up to $t=10^5$.
Dashed lines (slightly displaced for a better readability):
least-square fits of data for $t>10^3$.}
\label{numrad}
\end{center}
\end{figure}

Figure~\ref{numrad} shows numerical data
for the mean number $\mean{N_t}$ of radial records of Polya and Pearson planar walks,
plotted against $\sqrt{t}$ up to $t=10^5$.
Both datasets exhibit a very accurate square-root growth law.
The slopes $A$ of the least-square fits shown as dashed lines,
and the corresponding values of the distance $a$ according to~(\ref{anrad}),~(\ref{m1}),
are given in Table~\ref{arad} for both kinds of walks.

\begin{table}[!ht]
\begin{center}
\begin{tabular}{|l|c|c|}
\hline
walk & $A$ & $a$ \cr
\hline
Polya & 2.10 & 0.559 \cr
Pearson & 2.35 & 0.500 \cr
\hline
\end{tabular}
\caption
{Numerical values of amplitude $A$ of power-law growth~(\ref{nrad})
of mean number of radial records for Polya and Pearson planar walks,
as extracted from the data shown in Figure~\ref{numrad},
and corresponding values of distance $a$ according to~(\ref{anrad}),~(\ref{m1}).}
\label{arad}
\end{center}
\end{table}

\begin{figure}[!ht]
\begin{center}
\includegraphics[angle=0,width=.6\linewidth,clip=true]{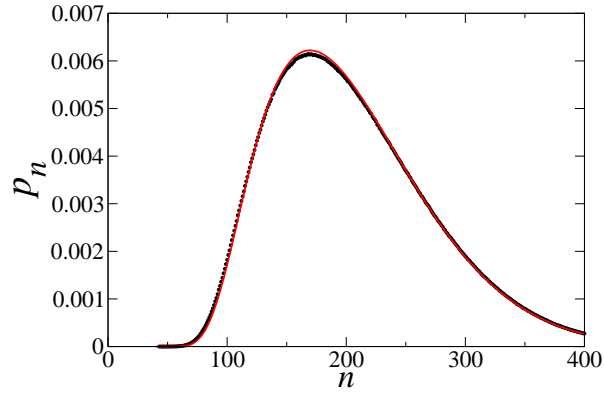}
\caption{\small
Full distribution of number of radial records, $p_n(t)=\prob{N_t=n}$, for planar Polya walks at time~$t=10^4$.
Black symbols: numerical data.
Full red curve: distribution~(\ref{fm}) of rescaled variable $U$ (see text).}
\label{hrad}
\end{center}
\end{figure}

Figure~\ref{hrad} shows the distribution of the number of radial records,
$p_n(t)=\prob{N_t=n}$, for Polya walks with $t=10^4$ steps.
Numerical data
(symbols)
are in very good agreement with the distribution~(\ref{fm}) of the rescaled variable~$U$ (full curve).
The constant of proportionality between $N_t$ and $U$
has been fixed by using the true finite-time mean record number $\mean{N_t}\approx205.2$.
This procedure is numerically more accurate than using the asymptotic growth law~(\ref{nrad}).
Corrections to scaling are again very small (of the order of one percent).
This plot provides a strong corroboration of the expected law of proportionality~(\ref{propor}).

For small values of $U$, the distribution $f_U(U)$ is dominated by the first term in~(\ref{fm}), i.e.,
\beq
f_U(U)\approx\frac{c}{U^3}\,\e^{-j_1^2/(4U^2)}.
\eeq
with $j_1\approx2.404825$ and $c=j_1/J_1(j_1)\approx4.632258$.
This exponentially small left tail is to be contrasted with the distribution of the number of records
for renewal processes (see~(\ref{nx}),~(\ref{fcontour})).
In that situation,
the distribution of the rescaled variable $X$ is non-zero ---and in fact maximal--- at $X=0$
for all $\theta<1/2$.

\subsection{Higher dimensions}
\label{radhi}

The general setting exposed in Section~\ref{radgen}
and illustrated in two dimensions in Section~\ref{rad2} remains valid in any dimension.
The aim of this section is to emphasize a few simplifying features at large dimensions.

First of all, as dimension $d$ increases,
the asymptotic distribution of the radius~$R_t$ becomes more and more narrow.
We have indeed (see~(\ref{smoms}))
\beq
\frac{\mean{S^2}}{\mean{S}^2}=\frac{d}{2}\left(\frac{\Gamma(d/2)}{\Gamma((d+1)/2)}\right)^2
=1+\frac{1}{2d}+\frac{1}{8d^2}+\cdots
\eeq
It is then clear that the distribution of the largest radius $\bar R_t$ has the same property.
In other words, as dimension~$d$ increases,
both rescaled variables $S$ and $U$ converge to the same deterministic value,
which is unity, as $\mean{S^2}=1$ in any dimension.
This reads formally
\beq
\lim_{d\to\infty}f_S(S)=\delta(S-1),\quad
\lim_{d\to\infty}f_U(U)=\delta(U-1).
\label{sucertain}
\eeq

\begin{figure}[!ht]
\begin{center}
\includegraphics[angle=0,width=.6\linewidth,clip=true]{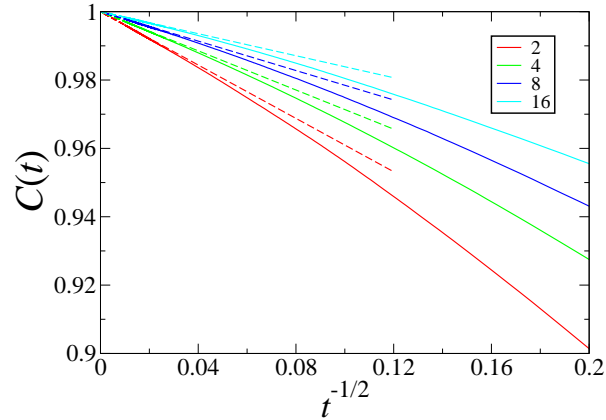}
\caption{\small
Angular correlation $C(t)$ between two most recent radial records
for Polya walks in dimensions $d=2$, 4, 8 and 16 (see legend)
against $t^{-1/2}$ up to $t=10^4$.
Full curves: numerical data.
Dashed lines: linear fits demonstrating asymptotic behavior~(\ref{cas}).}
\label{corrs}
\end{center}
\end{figure}

Let us now turn to the geometrical arrangement of radial records.
Figure~\ref{exemples} suggests that these records, shown as green symbols,
occur in long worm-like sequences of consecutive points,
interrupted by scarce non-local jumps.
In order to elaborate on this observation,
we introduce the angular correlation $C(t)$ between the two most recent records at time $t$, namely
\beq
C(t)=\bigmean{\cos\Theta\!\left(\vec R_{(N_t)},\vec R_{(N_t-1)}\right)}
=\bigmean{\frac{\vec R_{(N_t)}\cdot\vec R_{(N_t-1)}}{R_{(N_t)}\,R_{(N_t-1)}}}
\eeq
(conditioned on having $N_t\ge2$).
Figure~\ref{corrs} shows the angular correlation $C(t)$
for Polya walks in dimensions $d=2$, 4, 8 and 16 (see legend),
plotted against $t^{-1/2}$ up to $t=10^4$.
The dashed lines demonstrate a slow convergence of $C(t)$ to unity, of the form
\beq
C(t)\approx1-\frac{B}{\sqrt{t}}.
\label{cas}
\eeq
The amplitude $B$ of the leading correction is non-universal, i.e., depends on the kind of walk.
Both for Polya walks (Figure~\ref{corrs}) and for Pearson walks (not shown),
$B$ is observed to decay slowly to zero, roughly proportionally to $d^{-1/2}$, as dimension increases.

The following picture therefore emerges in the regime of large spatial dimensions.
Radial records occur in longer and longer worm-like sequences.
The effect of non-local jumps between these sequences becomes negligible in large dimensions,
as testified by the fall-off of the correction amplitude $B$.
The mean distance $a$ between the radii of successive records
can therefore be estimated by considering an effective one-dimensional problem.
The distances
\beq
r_n=X_{(n+1)}-X_{(n)}
\eeq
between successive records of one-dimensional random walks
have been investigated recently~\cite{gmsrev,gmsprl}.
They admit a non-trivial stationary distribution $f_r(r)$,
depending on the whole step distribution $f_x(x)$ defining the random walk.
For a symmetric continuous distribution such that $\mean{x^2}=2D$ is convergent,
their mean value $\mean{r}$ has a simple universal expression~\cite[ch.~XVIII.5]{feller2}:
\beq
\mean{r}=\sqrt D.
\label{spitzer}
\eeq
This result can be taken as a proxy for the distance $a$.
Using~(\ref{diff}), this reads
\beq
a\approx\frac{1}{\sqrt{2d}}.
\eeq
Inserting this estimate into~(\ref{anrad}),
and using $\mean{U}\approx1$ (see~(\ref{sucertain})),
we obtain the asymptotic expression
\beq
A\approx\sqrt{2d}
\label{a2d}
\eeq
for the amplitude $A$ of the growth law~(\ref{nrad}) of the mean number $\mean{N_t}$ of radial records.
This prediction is expected to hold to leading order as $d\gg1$ for all walks consisting of unit steps.

\begin{figure}[!ht]
\begin{center}
\includegraphics[angle=0,width=.6\linewidth,clip=true]{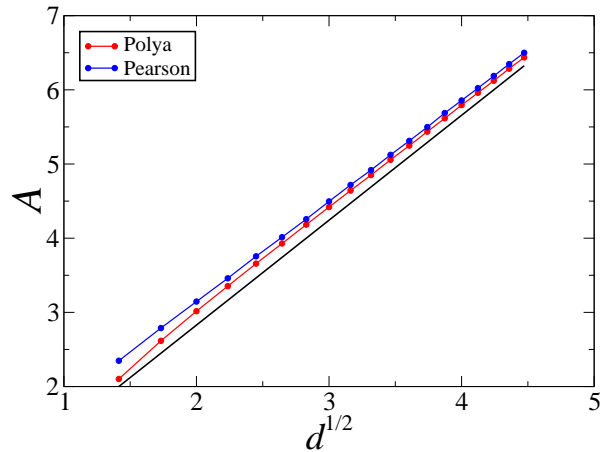}
\caption{\small
Symbols:
amplitudes $A$ of asymptotic growth law~(\ref{nrad})
of mean number of radial records for Polya and Pearson walks (see legend)
against $\sqrt{d}$ for all dimensions $d$ from 2 to 20.
Straight line with slope $\sqrt2$:
prediction~(\ref{a2d}) at large dimension.}
\label{amplis}
\end{center}
\end{figure}

Figure~\ref{amplis} shows the amplitudes $A$ of the growth law of the mean number $\mean{N_t}$
of radial records of Polya and Pearson walks in all dimensions up to $d=20$.
Just as in the two-dimensional situation (Figure~\ref{numrad}),
amplitudes have been extracted by means of least-square fits of data in the range $10^3<t<10^5$.
These amplitudes are plotted against $\sqrt{d}$.
Both datasets corroborate
the scaling law~(\ref{a2d}) at large dimension,
shown as a black straight line with slope $\sqrt2$.
The difference between the amplitudes $A$ for Pearson and Polya walks
is observed to decrease rapidly as a function of dimension and to reach
a non-zero limit $\Delta A\approx0.06$.
This observation suggests that the first correction to the prediction~(\ref{a2d})
is finite and non-universal.

\section{Discussion}
\label{disc}

In this paper we have investigated the statistics of various kinds of records
associated with planar random walks.
We have chosen three characteristic examples of records among many possibilities,
namely diagonal, simultaneous and radial records.
These examples are appealing and interesting in several regards.
Their definition is simple and natural, involving only elementary geometrical considerations.
The statistics of these records is varied.
The mean numbers of records of each kind grow as power laws of time,
with the simple rational exponents 1/4, 1/3 and 1/2.

The investigation of diagonal and simultaneous records of Polya walks,
performed in Sections~\ref{diag} and~\ref{simul},
relies upon the underlying renewal structure of the successive hitting times
and locations of translated copies of a fixed target.
In this sense, this work represents a two-dimensional extension of
the analysis made by Feller
of ladder points, i.e., records for one-dimensional random walks.
Renewal theory has allowed us to derive by analytical means a vast panoply of results at large times,
including the full statistics of the numbers of diagonal and simultaneous records (Section~\ref{diagfull}),
the joint law of the epoch and location of the current diagonal record (Section~\ref{diagel}),
and the angular distribution of the current simultaneous record (Section~\ref{simulang}).
All these asymptotic results are universal, at least among the class of walks consisting of unit steps,
whereas most of them extend to the whole class of isotropic diffusive walks.

Radial records of isotropic random walks in any spatial dimension $d$ have been investigated in Section~\ref{rad}.
This study is partly heuristic, and thus different in spirit from the previous ones,
for the mere reason that the sequence of radial records does not follow a renewal scheme.
The mean number of radial records has been shown to follow a superuniversal law,
growing as $\sqrt{t}$ irrespective of dimension.
The associated amplitude becomes itself universal at large spatial dimensions,
growing as $\sqrt{2d}$, irrespective of the kind of diffusive walk under consideration,
at least for walks consisting of unit steps.
The full statistics of the number of records has also been characterized.

The present work could be further extended in a great many directions.
As stressed in Section~\ref{intro}, a whole breadth of different records can be attached to a planar random walk
such as the Polya walk on the square lattice,
besides the three cases we have chosen to explore in detail.
Moreover, besides the isotropic diffusive walks considered in this work,
many other types of random walks could be considered a priori,
including anisotropic diffusive walks, walks possessing a drift velocity,
L\'evy walks whose step distribution has an infinite variance,
and various kinds of walks with correlated steps.

To close, let us discuss how our main findings concerning isotropic walks
are changed in higher spatial dimensions $d$.
The case of radial records has been studied in Section~\ref{rad} and summarized just above.
The statistics of diagonal and simultaneous records exhibits a more interesting dependence on $d$.
The definitions of both kinds of records extend in a straightforward way
to Polya walks in any dimension, e.g.~on hypercubic lattices,
as well as the reduction of the problem to renewal processes.
Now, let us remind the general fact that two objects of dimensions $d_1$ and~$d_2$
embedded in $d$-dimensional space
intersect easily if $d<d_1+d_2$, as their intersection is typically an object of dimension
$\delta=d_1+d_2-d$.\footnote{This formula is an extension to generic objects
of the well-known identity for linear subspaces
$\dim E+\dim F=\dim (E\cap F)+\dim (E+F)$.}
In the opposite case ($d>d_1+d_2$), the objects intersect either scarcely or not at all.
The situation where $\delta=0$, i.e., $d=d_1+d_2$, is the marginal one.
Within this setting, let us first consider diagonal records.
The target is the main diagonal of the lattice in the direction $(1,1,\dots,1)$.
The walk has $d_1=2$, as a consequence of diffusive scaling,
whereas the target has $d_2=1$, so that the marginal situation is $d=3$.
The corresponding survival probability,
which decays as $S(t)\sim t^{-1/4}$ for $d=2$,
can indeed be argued to decay as $S(t)\sim(\ln t)^{-1/2}$ for $d=3$,
and to reach a non-zero escape probability $S_\infty$ for $d\ge4$ (see~\cite{BKC,BKM,KR}).
As a consequence,
the mean number of records grows as $\mean{N_t}\sim1/S(t)\sim(\ln t)^{1/2}$ for $d=3$,
whereas for $d\ge4$ the total number $N_\infty$ of records of an infinitely long walk
is finite, and geometrically distributed with $\mean{N_\infty}=(1-S_\infty)/S_\infty$.
The case of simultaneous records is equally interesting.
There, the target is an orthant, or hyperoctant,
whose tip is initially located at the point $(1,1,\dots,1)$.
The walk still has $d_1=2$, whereas the target now has $d_2=d$,
so that both objects intersect easily in any dimension.
The corresponding survival probability indeed decays as a universal power law of the form
$S(t)\sim t^{-\theta(d)}$ in any spatial dimension $d$,
where the survival exponent~$\theta(d)$ only depends on $d$~\cite{BKC}.
We have $\theta(1)=1/2$, $\theta(2)=1/3$, $\theta(3)\approx0.22708$~\cite{bogo},
and $\theta(d)$ is known to become exponentially small for large~$d$~\cite{BKC,BKM}.
Thus, in any spatial dimension $d$, the mean number of records grows as $\mean{N_t}\sim t^{\theta(d)}$,
and its full distribution is universal and given by renewal theory.
Table~\ref{asys} summarizes the above discussion.

\begin{table}[!ht]
\begin{center}
\begin{tabular}{|l|c|c|c|}
\hline
Records & $d=2$ & $d=3$ & $d\ge4$ \cr
\hline
Diagonal & $t^{1/4}$ & $(\ln t)^{1/2}$ & finite \cr
Simultaneous & $t^{1/3}$ & $t^{\theta(3)}$ & $t^{\theta(d)}$ \cr
Radial & $t^{1/2}$ & $t^{1/2}$ & $t^{1/2}$ \cr
\hline
\end{tabular}
\caption
{Growth laws of numbers $N_t^{({\rm D})}$, $N_t^{({\rm S})}$, and $N_t^{({\rm R})}$
of diagonal, simultaneous, and radial records of isotropic random walks in dimension $d\ge2$.}
\label{asys}
\end{center}
\end{table}

\ack
It is a pleasure to thank Paul Krapivsky for very stimulating discussions
at various stages of this work.

\appendix

\section{Additive and multiplicative observables in renewal theory}
\label{apprenew}

This Appendix is devoted to the evaluation of additive and multiplicative observables
attached to a renewal process in continuous time.
Notations are consistent with those used in the body of the paper.
In particular, renewal events are referred to as records.
Within this setting,
many time-dependent quantities can be determined explicitly in Laplace space.
Hereafter we adopt the line of thought and the notations
of our earlier work~\cite{us}.
Let the temporal increments $\tau_n$ be iid variables
drawn from a continuous distribution with density $f_\tau(\tau)$.
For a given time $t$,
the number of records is the unique integer $N_t$ such that $T_{(N_t)}\le t<T_{(N_t+1)}$,
with the definition
\beq
T_{(n)}=\tau_1+\cdots+\tau_n.
\eeq
The number $N_t$ of records is random, as it depends on the whole renewal process $\{\tau_n\}$.

\subsection*{Additive observables}

An additive observable is a quantity of the form
\beq
A(t)=\bigmean{\sum_{n=1}^{N_t}a(\tau_n)},
\label{adef}
\eeq
where $a(\tau)$ is an arbitrary given function of the temporal increment $\tau$.
The definition of $a(\tau)$ may involve averaging over other random variables,
as long as they are statistically independent of $\tau$.
A renewal process endowed with an additive observable of this kind
is referred to as a renewal reward process~\cite{cox}.
Continuous-time random walks~\cite{montroll} belong to this class of processes.

The quantity $A(t)$ can be evaluated as follows.
We have
\beq
A(t)=\sum_{N=0}^\infty\bigmean{\sum_{n=1}^Na(\tau_n)\,\vec{1}(T_{(N)}<t<T_{(N)}+\tau_{N+1})},
\eeq
where $N$ is a shorthand for $N_t$
and $\vec{1}(\cdot)$ denotes the indicator function of an event.
In Laplace space, this translates to
\beq
\h A(s)=\sum_{N=0}^\infty\bigmean{\sum_{n=1}^Na(\tau_n)
\,\e^{-sT_{(N)}}\frac{1-\e^{-s\tau_{N+1}}}{s}}.
\eeq
Averages over the iid $\tau$ variables boil down to two simple integrals, i.e.,
\beq
\h f_\tau(s)=\int_0^\infty f_\tau(\tau)\e^{-s\tau}\,\dd\tau,\quad
g_A(s)=\int_0^\infty f_\tau(\tau)a(\tau)\e^{-s\tau}\,\dd\tau.
\eeq
We thus obtain
\beq
\h A(s)=\frac{g_A(s)}{s(1-\h f_\tau(s))}.
\label{ares}
\eeq

The simplest of all additive observables, corresponding to the choice
\beq
a(\tau)=1,
\eeq
yields $A(t)=\mean{N_t}$,
the mean number of records at time $t$.
We have then $g_A(s)=\h f_\tau(s)$, so that~(\ref{ares}) becomes~(\ref{navelap}), as should~be.

Consider now the power-law observable
\beq
a(\tau)=\tau^\beta,
\eeq
and a distribution $f_\tau(\tau)$ with a power-law tail of the form
\beq
f_\tau(\tau)\approx\frac{c}{\tau^{\theta+1}}.
\eeq
The range of exponents of interest for the present purpose is
\beq
0<\theta<1,\quad\beta>\theta.
\label{ssrange}
\eeq
For instance,
diagonal and simultaneous records respectively correspond to $\theta=1/4$ and $\theta=1/3$,
whereas the exponent $\beta=1/2$ dictated by diffusive scaling
provides a toy model for the abscissa of the current record.
All over the parameter range~(\ref{ssrange}), we obtain the estimates
\beq
1-\h f_\tau(s)\approx\frac{\Gamma(1-\theta)}{\theta}\,c\,s^\theta,\quad
g_A(s)\approx\Gamma(\beta-\theta)c\,s^{\theta-\beta},
\label{fgasy}
\eeq
for $s\to0$,
so that
\beq
\h A(s)\approx\frac{\theta\Gamma(\beta-\theta)}{\Gamma(1-\theta)}\,s^{-\beta-1}.
\eeq
We thus predict a power-law growth of the form
\beq
A(t)\approx\frac{\theta\Gamma(\beta-\theta)}{\Gamma(1-\theta)\Gamma(\beta+1)}\,t^\beta.
\label{apower}
\eeq
This result is universal,
in the sense that it only involves the exponents $\theta$ and~$\beta$.

Non-universal asymptotic results show up
outside the self-similar range~(\ref{ssrange}).
Let us just give one example.
For $0<\beta<\theta<1$, the estimate~(\ref{fgasy}) for $g_A(s)$
is to be replaced by the non-universal constant
\beq
g_A(0)=\overline{a}=\int_0^\infty f_\tau(\tau)a(\tau)\dd\tau,
\eeq
so that we have
\beq
A(t)\approx\overline{a}\,\mean{N_t}\approx\frac{\overline{a}}{c}\,\frac{\sin\pi\theta}{\pi}\,t^\theta.
\eeq
At variance with~(\ref{apower}),
this result is non-universal,
as it involves the ratio of two microscopic constants $\overline{a}$ and $c$.

\subsection*{Multiplicative observables}

A multiplicative observable is a quantity of the form
\beq
B(t)=\bigmean{\prod_{n=1}^{N_t}b(\tau_n)},
\label{bdef}
\eeq
where $b(\tau)$ is an arbitrary given function of the temporal increment $\tau$.
Here again, the definition of $b(\tau)$ may involve averaging over other random variables,
as long as they are statistically independent of $\tau$.

The quantity $B(t)$ can be evaluated as follows.
We have
\beq
B(t)=\sum_{N=0}^\infty\bigmean{\prod_{n=1}^Nb(\tau_n)\,\vec{1}(T_{(N)}<t<T_{(N)}+\tau_{N+1})}.
\eeq
In Laplace space, this translates to
\beq
\h B(s)=\sum_{N=0}^\infty\bigmean{\prod_{n=1}^Nb(\tau_n)
\,\e^{-sT_{(N)}}\frac{1-\e^{-s\tau_{N+1}}}{s}}.
\eeq
Averages over the iid $\tau$ variables again boil down to two simple integrals,
namely $\h f_\tau(s)$ and
\beq
h_B(s)=\int_0^\infty f_\tau(\tau)b(\tau)\e^{-s\tau}\,\dd\tau.
\eeq
We thus obtain
\beq
\h B(s)=\frac{1-\h f_\tau(s)}{s(1-h_B(s))}.
\label{bres}
\eeq

The simplest of all multiplicative observables,
corresponding to the choice
\beq
b(\tau)=z,
\eeq
where $z$ is an arbitrary constant,
yields
\beq
B(t)=\mean{z^{N_t}}=\sum_{n=0}^\infty p_n(t)z^n,
\eeq
the generating function of the probabilities $p_n(t)$.
We have then $h_B(s)=z\h f_\tau(s)$, so that~(\ref{bres}) reads
\beq
\h B(s)=\frac{1-\h f_\tau(s)}{s(1-z\h f_\tau(s))}.
\eeq
The expression~(\ref{nlap}) is recovered by expanding the above result
as a power series in~$z$, as should~be.

\section{Derivation of Equation~(\ref{bsca})}
\label{appderiv}

This Appendix is devoted to the derivation of~(\ref{bsca})
for an arbitrary survival exponent in the range $0<\theta<1/2$.
The essential part of the derivation consists in evaluating the expression (see~(\ref{bdeno}))
\beq
1-h_B(p,s)\approx-\int_0^\infty S(\tau)\dd\tau\,
\frac{\dd}{\dd\tau}\left(\e^{-s\tau}
\int_0^\infty f_\xi(\xi)\e^{-p\sqrt\tau\xi}\,\dd\xi\right).
\eeq
Using the power-law tail~(\ref{sasy}) of $S(\tau)$
and the distribution~(\ref{fxires}) of~$\xi$, this reads
\beqa
1-h_B(p,s)&\approx&\frac{2c}{\Gamma(\theta+1)}\int_0^\infty\tau^{-\theta}\,\e^{-s\tau}\,\dd\tau
\nonumber\\
&\times&
\int_0^\infty\xi^{2\theta-1}\,\e^{-p\sqrt\tau\xi-\xi^2}\left(s+\frac{p\xi}{2\sqrt{\tau}}\right)\dd\xi.
\eeqa
Changing variables from $\tau$ to $u=s\tau$ and from $\xi$ to $v=\xi\sqrt{u}$,
and introducing the dimensionless variable $z=p/\sqrt{s}$,
we obtain
\beqa
1-h_B(p,s)&\approx&\frac{2c\,s^\theta}{\Gamma(\theta+1)}
\int_0^\infty u^{-2\theta}\,\e^{-u}\,\dd u
\nonumber\\
&\times&
\int_0^\infty v^{2\theta-1}\,\e^{-zv-v^2/u}\left(1+\frac{zv}{2u}\right)\dd v.
\eeqa
The integration over $u$ can be performed first, yielding
\beq
1-h_B(p,s)\approx\frac{2c\,s^\theta}{\Gamma(\theta+1)}
\int_0^\infty\e^{-zv}(2K_{1-2\theta}(2v)+z K_{-2\theta}(2v))\dd v,
\eeq
where $K_\nu$ is the modified Bessel function.
The integration over $v$ can also be worked out.
We thus obtain
\beq
1-h_B(p,s)\approx\frac{2c\,s^\theta}{\Gamma(\theta+1)}\,\frac{\pi\cos2\theta\gamma}{\sin2\pi\theta},
\label{denosca}
\eeq
with
\beq
z=\frac{p}{\sqrt{s}}=2\cos\gamma.
\label{zsca}
\eeq

Inserting the estimates~(\ref{laps}) and~(\ref{denosca}) into~(\ref{bfullres}), we obtain
\beq
\h B(\sigma,p,s)\approx\frac{(1+y)^{-\theta}}{s\,\phi((1+y)^{-1/2}z)},
\label{appbsca}
\eeq
where the scaling variables $y$ and $z$ read
\beq
y=\frac{\sigma}{s},\quad z=\frac{p}{\sqrt{s}},
\eeq
and the scaling function $\phi(\zeta)$ is given in parametric form by
\beq
\phi=\frac{\cos 2\theta\gamma}{\cos\pi\theta},\quad
\zeta=2\cos\gamma.
\label{appphi}
\eeq
The point $\zeta=0$ corresponds to $\gamma=\pi/2$, so that $\phi(0)=1$, as should be.

Whenever $\theta$ is rational,
the scaling function $\phi(\zeta)$ is an algebraic function.
More precisely, for $\theta=p/q$ in irreducible form, with $p<q/2$ (since $0<\theta<1/2$),
the algebraic degree of $\phi$ is~$q$ (if $q$ is odd) or $q/2$ (if $q$ is even).

There is one single case in degree 2.
This is $\theta=1/4$, corresponding to diagonal records investigated in Section~\ref{diag}.
We have
\beq
\zeta=2(\phi^2-1),
\eeq
hence the explicit expression
\beq
\phi(\zeta)=\left(1+\frac{\zeta}{2}\right)^{1/2}.
\label{appphi4}
\eeq

Two cases pertain to degree 3.
The first one is $\theta=1/3$,
corresponding to simultaneous records investigated in Section~\ref{simul},
where we obtain the equation of a cubic curve:
\beq
\zeta^2=(\phi-1)^2(\phi+2).
\label{appphi3}
\eeq
This curve is said to be unicursal, as it admits the rational parametrization
\beq
\phi=u^2-2,\quad\zeta=u(u^2-3).
\label{rat}
\eeq
The second case in degree 3 is $\theta=1/6$,
where we have
\beq
\zeta=3\sqrt{3}\phi(\phi^2-1).
\eeq

It is remarkable that the two situations met in Sections~\ref{diag} and~\ref{simul}
of the body of this paper are among the first three cases of the above classification.

\section{Law of the ratio of two correlated random variables}
\label{appratio}

In this Appendix we investigate the distribution $f_\lam(\lam)$ of the ratio
\beq
\lam=\frac{Y}{X}
\label{lamdef}
\eeq
of two positive random variables with an arbitrary joint distribution.

\noindent $\bullet$ Consider first the case where the density $f_{X,Y}(X,Y)$
of the joint distribution is known.
We have then
\beqa
f_\lam(\lam)&=&\int_0^\infty\dd X\int_0^\infty f_{X,Y}(X,Y)\,\delta\!\left(\lam-\frac{Y}{X}\right)\dd Y
\nonumber\\
&=&\int_0^\infty X\,f_{X,Y}(X,\lam X)\dd X.
\eeqa

\noindent $\bullet$ Consider now the case where only the bivariate characteristic function
\beq
B(p,q)=\bigmean{\e^{-pX-qY}}
\eeq
of the joint distribution is known.
We have then
\beq
f_{X,Y}(X,Y)=\int\frac{\dd p}{2\pi\ii}\int\frac{\dd q}{2\pi\ii}\,B(p,q)\e^{pX+qY},
\eeq
and so (formally)
\beq
f_\lam(\lam)=\int\frac{\dd p}{2\pi\ii}\int\frac{\dd q}{2\pi\ii}\,B(p,q)
\int_0^\infty X\,\e^{(p+\lam q)X}\,\dd X.
\eeq
Let us assume for a while that the density $f_{X,Y}(X,Y)$ falls off exponentially in both variables,
so that $B(p,q)$ is analytic when $\Re p$ and $\Re q$ are both larger than $-a$ for some positive~$a$.
We can therefore choose $\Re p$ and $\Re q$ to be small and negative.
The integral over $X$ then reads $1/(p+\lam q)^2$,
whereas the integral over $p$ is given by the residue at the double pole at $p=-\lam q$.
We thus obtain
\beq
f_\lam(\lam)=-\int\frac{\dd q}{2\pi\ii}\left.\frac{\partial B(p,q)}{\partial p}\right|_{p=-\lam q}.
\label{fratio}
\eeq
The above expression makes sense for an arbitrary joint distribution $f_{X,Y}(X,Y)$.
The integration contour can indeed be chosen to be the imaginary axis,
so that~(\ref{fratio}) only involves the bivariate Fourier transform of $f_{X,Y}(X,Y)$.

Whenever the variables $X$ and $Y$ are exchangeable,
i.e., $f_{X,Y}(X,Y)=f_{X,Y}(Y,X)$ or $B(p,q)=B(q,p)$,
the product
\beq
\lam f_\lam(\lam)=\frac{1}{\lam}\,f_\lam\!\left(\frac{1}{\lam}\right)
\label{lamflam}
\eeq
is invariant under the change of $\lam$ into its inverse.

Let us illustrate the above by considering the example
where $X$ and $Y$ are two identical independent L\'evy stable variables
with index in the range $0<\theta<1$ and a suitable chosen scale factor, such that
\beq
B(p,q)=\e^{-p^\theta-q^\theta},
\eeq
Equation~(\ref{fratio}) then reads
\beq
f_\lam(\lam)=\theta\int\frac{\dd q}{2\pi\ii}\,(-\lam q)^{\theta-1}\,\e^{-q^\theta-(-\lam q)^\theta}.
\eeq
Setting $q=\ii y$ and dealing separately with the ranges $y>0$ and $y<0$,
some algebra leads us to the expression
\beq
f_\lam(\lam)=\frac{\sin\pi\theta}{\pi\lam}\,\frac{1}{\lam^\theta+\lam^{-\theta}+2\cos\pi\theta}.
\eeq
We have thus recovered the celebrated Lamperti law~\cite{lamperti}.
The above expression obeys the symmetry~(\ref{lamflam}), as should be.

\section*{References}

\bibliography{paper.bib}

\end{document}